\documentclass[prd,twocolumn,showpacs,reprint,preprintnumbers,nofootinbib]{revtex4-1}

\RequirePackage[colorlinks=true
,urlcolor=blue
,anchorcolor=blue
,citecolor=blue
,filecolor=blue
,linkcolor=blue
,menucolor=blue
,pagecolor=blue
,linktocpage=true
,pdfproducer=medialab
,pdfa=true
]{hyperref}
\usepackage{hyperref}

\usepackage{amsmath,amssymb,amsthm,amsfonts}
\usepackage{graphicx,tabularx}
\usepackage{color}
\usepackage{multirow}  
\usepackage{comment}
\usepackage{mathrsfs}
\usepackage{slashed}

\makeatletter
\def\l@subsubsection#1#2{}
\makeatother

\newcommand{\be}{\begin{equation}\begin{aligned}}
\newcommand{\ee}{\end{aligned}\end{equation}}

\newcommand{\mm}{\text{mm}}
\newcommand{\cm}{\text{cm}}
\newcommand{\m}{\text{m}}

\newcommand{\mrad}{\text{mrad}}
\newcommand{\ns}{\text{ns}}

\newcommand{\mev}{\text{MeV}}

\newcommand{\gev}{\text{GeV}}
\newcommand{\tev}{\text{TeV}}
\newcommand{\iab}{\text{ab}^{-1}}
\newcommand{\ifb}{\text{fb}^{-1}}

\newcommand{\met}{\slashed{\text{E}}_\text{T}}
\newcommand{\mdm}{m_{_\text{DM}}}

\renewcommand{\eqref}[1]{Eq.~(\ref{eq:#1})}
\newcommand{\secref}[1]{Sec.~\ref{sec:#1}}
\newcommand{\figref}[1]{Fig.~\ref{fig:#1}}

\newcommand{\appref}[1]{Appendix~\ref{sec:#1}}

\newcommand{\nl}{\nonumber \\}
\newcommand{\x}{\chi}
\newcommand{\Ap}{A^\prime}
\newcommand{\mAp}{m_{A^\prime}}
\newcommand{\eps}{\epsilon}
\newcommand{\Lag}{\mathscr{L}}
\newcommand{\order}[1]{\mathcal{O}{(#1)}}

\newcommand{\MeV}{\text{ MeV}}
\newcommand{\GeV}{\text{ GeV}}
\newcommand{\TeV}{\text{ TeV}}

\begin{document}
%\title{The Inelastic Frontier at the LHC}
\title{Inelastic Dark Matter at the LHC Lifetime Frontier: \\ ATLAS, CMS, LHCb, CODEX-b, FASER, and MATHUSLA}

\preprint{SLAC-PUB-17325}
\preprint{UCI-TR-2018-08}

\author{Asher Berlin}
\email{berlin@slac.stanford.edu}
\affiliation{SLAC National Accelerator Laboratory, 2575 Sand Hill Road, Menlo Park, CA, 94025, USA}

\author{Felix Kling}
\email{fkling@uci.edu}
\affiliation{Department of Physics and Astronomy, University of
California, Irvine, CA 92697, USA}

\begin{abstract}
Visible signals from the decays of light long-lived hidden sector particles have been extensively searched for at beam dump, fixed-target, and collider experiments. If such hidden sectors couple to the Standard Model through mediators heavier than $\sim 10$ GeV, their production at low-energy accelerators is kinematically suppressed, leaving open significant pockets of viable parameter space. We investigate this scenario in models of inelastic dark matter, which give rise to visible signals at various existing and proposed LHC experiments, such as ATLAS, CMS, LHCb, CODEX-b, FASER, and MATHUSLA. These experiments can leverage the large center of mass energy of the LHC to produce GeV-scale dark matter from the decays of dark photons in the cosmologically motivated mass range of $\sim 1-100$ GeV. We also provide a detailed calculation of the radiative dark matter-nucleon/electron elastic scattering cross section, which is relevant for estimating rates at direct detection experiments. 
\end{abstract}
\date{\today}
\maketitle
%\tableofcontents

%%%%%%%%%%%%%%%%%%%%%%%%%%%%%%%%%%%
%%% Introduction                           
%%%%%%%%%%%%%%%%%%%%%%%%%%%%%%%%%%%
\section{Introduction}
\label{sec:introduction}

In recent years, an extensive program has emerged in the search for GeV-scale hidden sectors, involving beam dump, fixed-target, and collider experiments~\cite{Alexander:2016aln,Battaglieri:2017aum}. If the mediators coupling such hidden sectors to the Standard Model (SM) are heavier than roughly $\sim 10 \GeV$ (corresponding to the center of mass energy of $B$-factories), their production at low-energy accelerators is kinematically suppressed. As a result, there are large regions of viable parameter space for GeV-scale particles that couple to the SM at the $\order{10^{-2}}$-level through $\order{10} \GeV$ mediators.

These scenarios are directly motivated from considerations of light dark matter (DM). It is well-known that models of thermal DM at or below the GeV scale often require the presence of new light mediators in order to facilitate the depletion of the DM energy density in the early universe~\cite{Lee:1977ua}. Although stated less often, SM measurements (e.g., of the invisible widths of the $Z$ and Higgs bosons) extend these claims to masses as large as $\mdm \sim \text{few} \times 10 \GeV$~(see, e.g., Refs.~\cite{Escudero:2016gzx,Kearney:2016rng}). 

For $\mdm \lesssim 50 \GeV$, measurements of the cosmic microwave background (CMB) severely restrict how these new mediators can couple DM to the SM~\cite{Ade:2015xua}. For instance, considerations of the CMB necessitate that the DM annihilation rate to electromagnetically charged particles is suppressed at late times. This requirement typically also leads to small annihilation rates today, worsening the discovery potential for indirect detection experiments searching for the products of local DM annihilations. In this work, we focus on a class of models that naturally leads to negligible annihilation rates at the time of recombination. 

Although originally motivated as an explanation for the longstanding DAMA anomaly~\cite{TuckerSmith:2001hy}, models of inelastic DM (iDM) constitute a viable and compelling paradigm for light thermal DM. In these models, DM couples to the SM only by interacting with a nearly degenerate excited state. For relative mass-splittings larger than $\sim \order{10^{-6}}$, DM-nucleon/electron scattering at direct detection experiments is kinematically suppressed by the small DM virial velocity, $v \sim \order{10^{-3}}$. However, this kinematic suppression is overcome in relativistic settings, such as accelerators, where the DM and excited state can be directly produced. For mass-splittings above a few MeV, the excited state can decay back to DM and a pair of SM fermions, often on collider timescales. In this case, searches for visible displaced vertices at colliders and fixed-target experiments are sensitive to cosmologically motivated masses and couplings.

To date, the majority of existing studies on visible signals of iDM have focused on mediator masses at or below the GeV scale (see, however, Ref.~\cite{Izaguirre:2015zva}). In this work, we investigate mediators heavier than $\sim 10 \GeV$, where most high-intensity accelerators have little sensitivity due to their limited center of mass energy. This large open region of viable parameter space is well-suited for high-energy machines, such as the LHC. In recent years, different search strategies have been proposed for detecting long-lived particles (LLPs) at the existing LHC experiments, ATLAS, CMS, and LHCb. Furthermore, various dedicated experiments have been suggested, such as CODEX-b, FASER, and MATHUSLA. In this paper, we focus on the potential sensitivity of these existing and proposed experiments to visible signals of iDM near the GeV scale.

The remainder of this paper is structured as follows. In \secref{below_weakscale}, we motivate the existence of additional mediators for DM below the weak scale. In \secref{idm}, we introduce the particular iDM model that we investigate throughout this work and discuss various aspects of its cosmology in \secref{cosmo}. In \secref{constraints}, we summarize existing constraints on these models. In \secref{searches}, we present strategies for the LHC to search for displaced lepton-jets and time-delayed tracks at ATLAS and CMS and displaced vertices at LHCb, CODEX-b, FASER, and MATHUSLA. In \secref{prod}, we discuss details related to our modeling of iDM production at the LHC and present our results in \secref{results}. We summarize our conclusions in \secref{conclusion}. Detailed discussions of experimental energy thresholds, kinetically-mixed dark photons, loop-induced DM-nucleon/electron elastic scattering cross sections, and dark photon production via vector meson mixing are provided in Appendices~\ref{sec:threshold}, \ref{sec:mixing}, \ref{sec:DD}, and \ref{sec:VectorMesonMixing}, respectively.

%%%%%%%%%%%%%%%%%%%%%%%%%%%%%%%%%%%
%%% LOW MASS THERMAL DM                
%%%%%%%%%%%%%%%%%%%%%%%%%%%%%%%%%%%

\section{Thermal Dark Matter Below the Weak Scale}
\label{sec:below_weakscale}

Thermal DM that couples to the SM solely through the electroweak force must be heavier than the GeV scale. This is most often stated in terms of the Lee-Weinberg bound~\cite{Lee:1977ua}, which in its modern form relates the DM mass to the weak scale ($m_Z \sim 100 \GeV$), the temperature at matter-radiation equality ($T_\text{eq} \sim \text{eV}$), and the Planck mass ($m_\text{Pl} \sim 10^{18} \GeV$), 
\be
\label{eq:LeeWeinberg}
\mdm \gtrsim \frac{m_Z^2}{(T_\text{eq} \, m_\text{Pl})^{1/2}} \sim \text{GeV}
.
\ee
Eq.~(\ref{eq:LeeWeinberg}) implies that sub-GeV thermal DM motivates the presence of new light mediators~\cite{Boehm:2003hm}. This philosophy has guided various experimental developments in the search for well-defined ``thermal targets"~\cite{Alexander:2016aln,Battaglieri:2017aum,Berlin:2018bsc}.

This picture can be sharpened with the inclusion of additional experimental inputs. It is well-known that thermal DM that is lighter than $\sim 50 \GeV$ and annihilates through the exchange of an electroweak boson is strongly constrained by null results of recent direct detection experiments as well as by the invisible $Z$ and Higgs widths~\cite{Escudero:2016gzx,Kearney:2016rng}. This can be stated parametrically, similar to Eq.~(\ref{eq:LeeWeinberg}). If $\text{few} \times \mdm \lesssim m_Z$, then the DM annihilation rate ($\sigma v$) is smaller than the value required by thermal freeze-out ($\sigma v_\text{th} \sim 1 / T_\text{eq} \, m_\text{Pl}$),
\be
\frac{\sigma v}{\sigma v_\text{th}} \lesssim \order{10^{-2}} \times \frac{\Gamma_Z^\text{inv} \, T_\text{eq} \, m_\text{Pl}}{m_Z^3} \lesssim \order{10^{-2}}
~,
\ee
where $\Gamma_Z^\text{inv}$ is the additional contribution to the invisible $Z$ width from decays to DM pairs, and in the second inequality we have enforced $\Gamma_Z^\text{inv} \lesssim \order{\mev}$~\cite{Patrignani:2016xqp}. A similar statement holds for couplings to the SM Higgs. Such claims can be avoided, for instance, if $2 \, \mdm \simeq m_{Z,h}$, in which case invisible $Z$ or $h$ decays to DM particles are kinematically suppressed and DM annihilations are resonantly enhanced. Therefore, aside from this tuning, DM lighter than $\sim 50 \GeV$ motivates the presence of additional light mediators that are nearby in mass to the DM itself. A well-motivated example is the dark photon, as discussed in Sec.~\ref{sec:idm}. In the next section, we also introduce a class of models in which the DM-dark photon coupling is directly responsible for thermal freeze-out in the early universe. 

%%%%%%%%%%%%%%%%%%%%%%%%%%%%%%%%%%%
%%% iDM                            
%%%%%%%%%%%%%%%%%%%%%%%%%%%%%%%%%%%
\section{Inelastic Dark Matter}
\label{sec:idm}

Models of inelastic DM (iDM) were first proposed to resolve the longstanding DAMA anomaly~\cite{TuckerSmith:2001hy}. Although such explanations are no longer viable in light of searches conducted by more recent direct detection experiments (see, e.g., Ref.~\cite{Aprile:2018dbl}), the overall framework still constitutes a compelling paradigm for light thermal DM. 

In this section, we outline a concrete model that we will investigate throughout this work. We focus on a particular implementation of iDM, involving a Dirac pair of two-component Weyl fermions ($\eta$ and $\xi$) that are oppositely charged under a broken $U(1)_D$ symmetry ($Q_{\eta , \xi} = \pm 1$). The corresponding massive gauge boson is the dark photon, denoted as $\Ap$. The dark photon couples to SM hypercharge through the kinetic mixing term
\be
\label{eq:kinmix}
\Lag \supset \frac{\eps}{2 \cos{\theta_w}} \, \Ap_{\mu \nu} \, B^{\mu \nu}
~,
\ee
where $\eps \ll 1$ is a dimensionless parameter controlling the size of the mixing and $\theta_w$ is the Weinberg angle~\cite{Holdom:1985ag,delAguila:1988jz}. For example, this kinetic mixing is generated radiatively if there exist any particles charged under both $U(1)_D$ and hypercharge. In this case, the natural expected size is $\eps \sim e \, e_D / 16 \pi^2 \sim 10^{-3}$, where $e$ and $e_D$ are the electromagnetic and $U(1)_D$ gauge couplings, respectively. We remain agnostic concerning the origin of the dark photon mass, $\mAp$. If $\mAp \ll m_Z$, the $\Ap$ predominantly mixes with the SM photon and inherits a small coupling to SM fermions proportional to their electric charge, $\eps \, e \, Q_f$. For $\mAp \gtrsim \order{10} \GeV$, the $\Ap$ also significantly mixes with the SM $Z$. A detailed discussion of kinetic mixing for general $\Ap$ masses is provided in Appendix~\ref{sec:mixing}.

A Dirac mass, $m_D$, involving the DM spinors, $\eta$ and $\xi$, is allowed by all symmetries of the theory. In the case that a dark Higgs is responsible for the mass of the $\Ap$, it is natural to imagine that the spontaneous breaking of $U(1)_D$ also gives rise to Majorana mass terms ($\delta_{\eta, \xi}$) for these hidden sector fermions. At energies below the scale of $U(1)_D$-breaking, we parametrize the effective Lagrangian as
\be
\label{eq:IDMLag1}
- \Lag \supset m_D \, \eta \, \xi + \frac{1}{2} \, \delta_\eta \, \eta^2 + \frac{1}{2} \, \delta_\xi \, \xi^2 + {\text{h.c.}}
~,
 \ee
where $m_D , \delta_\eta , \delta_\xi > 0$. Since $\delta_{\eta , \xi} \neq 0$ explicitly breaks $U(1)_D$, it is technically natural to take $\delta_{\eta , \xi} \ll m_D$. In this limit, the mass eigenstates correspond to a pseudo-Dirac pair given by
\begin{align}
\x_1 &\simeq \frac{i}{\sqrt{2}} ~ (\eta - \xi) 
\nl
\x_2 &\simeq \frac{1}{\sqrt{2}} ~ (\eta + \xi)
~,
\end{align}
where the phase of $\x_1$ is fixed to guarantee a positive mass term. $\x_1$ and $\x_2$ are nearly degenerate, with masses $m_1 \lesssim m_2$ given by
\be
m_{1,2} \simeq m_D \mp \frac{1}{2} (\delta_\eta+\delta_\xi)
~.
\ee
For later convenience, we define the fractional mass-splitting between $\x_1$ and $\x_2$, 
\be
\Delta\equiv \frac{m_2-m_1}{m_1} \simeq \frac{\delta_\eta + \delta_\xi}{m_D} \ll 1
\, .
\ee

In four-component notation, $\x_{1,2}$ are described by Majorana spinors. In the pseudo-Dirac limit, they couple off-diagonally to the dark photon, i.e.,
\be
\label{eq:IDMLag2}
\Lag \supset ie_D ~ A^\prime_\mu ~ \bar\x_1\gamma^\mu\x_2 + \order{\delta_{\eta , \xi} / m_D}
~,
\ee
where $e_D$ is the $U(1)_D$ gauge coupling. In general, there also exist terms that couple $\Ap$ diagonally to $\x_1$ and $\x_2$. However, these are either suppressed by the mass hierarchy, $\delta_{\eta , \xi} / m_D \ll 1$, or vanish exactly when $\delta_\eta = \delta_\xi$.\footnote{This is due to an enhanced charge-conjugation symmetry under which $\Ap \to - \Ap$ and $\x_{1,2} \to \mp \x_{1,2}$.} Throughout this work, we will ignore such contributions, and assume that $\Ap$ couples purely off-diagonally (inelastically) to $\x_1 \x_2$ pairs.

%%%%%%%%%%%%%%%%%%%%%%%%%%%%%%%%%%%
%%% COSMO 
%%%%%%%%%%%%%%%%%%%%%%%%%%%%%%%%%%%

\section{Cosmology}
\label{sec:cosmo}

Kinetic mixing between $U(1)_D$ and hypercharge couples $\x_{1,2}$ to the SM. In particular, if the degree of kinetic mixing satisfies 
\be
\eps \gtrsim \order{10^{-8}} \times \left(\frac{\mAp}{\text{GeV}}\right)^{1/2}
~,
\ee
then the $\Ap$ equilibrates with the visible sector in the early universe. In this case, $\x_{1,2} - \Ap$ interactions often lead to a large thermal $\x_{1,2}$ number density that must be depleted at later times in order to be cosmologically viable. Since $m_2 \gtrsim m_1$, the depletion of $\x_2$ in the early universe can efficiently proceed through dynamics within the hidden sector, e.g., through scattering ($\x_2 \x_2 \to \x_1 \x_1$) or decays ($\x_2 \to \x_1 + \cdots$). $\x_1$ and $\x_2$ are odd under a global $\mathbb{Z}_2$ subgroup of $U(1)_D$, which is preserved by the Lagrangian terms of Eqs.~(\ref{eq:IDMLag1}) and (\ref{eq:IDMLag2}). Since $\x_1$ is the lightest state charged under this symmetry, it is stable on cosmological timescales and plays the role of DM. 

If $m_1 \gtrsim \mAp$, then DM freeze-out is dictated by annihilations to pairs of on-shell dark photons ($\x_1 \x_1 \to \Ap \Ap$), followed by $\Ap \to e^+ e^-, \mu^+ \mu^- , \cdots$. Such processes are unsuppressed at late times and for $m_1 \lesssim \text{few} \times 10 \GeV$ are severely constrained from considerations of the CMB and various indirect detection searches (see Ref.~\cite{Leane:2018kjk} for a recent review). If instead, $\mAp \gtrsim m_{1,2}$,  then freeze-out predominantly proceeds through coannihilations directly into SM particles, $\x_1 \x_2 \to A^{\prime \, *} \to f \bar{f}$, where $f$ is a SM fermion. For $m_f \ll m_1 \simeq m_2 \ll \mAp$, the thermally-averaged annihilation rate for this process is parametrically of the form
\be
\label{eq:sigmav1}
\langle \sigma v \rangle \sim \order{10^2} ~ \frac{\alpha_D \, \alpha_\text{em} \, \eps^2 \, m_1^2}{\mAp^4} ~ e^{- \Delta \, x}
~,
\ee
where $\alpha_D \equiv e_D^2 / 4 \pi$, $x \equiv m_1 / T$, and $T$ is the temperature of the photon bath. The number density of $\x_2$ is Boltzmann suppressed at late times. As a result, the coannihilation rate in Eq.~(\ref{eq:sigmav1}) is exponentially small at temperatures much below the $\x_1-\x_2$ mass-splitting ($\Delta \, x \gg 1$). 

As we will see below, for $\mAp \gtrsim 10 \GeV$, the degree of kinetic mixing is restricted by terrestrial experiments to be $\eps \lesssim \text{few} \times 10^{-2}$. Imposing this limit on $\eps$ and demanding that Eq.~(\ref{eq:sigmav1}) is of the required size for thermal freeze-out, i.e., $\langle \sigma v \rangle \sim \langle \sigma v_\text{th} \rangle \sim 1 / T_\text{eq} \, m_\text{Pl}$, then implies an upper bound on the mass-splitting,
\be
\Delta \lesssim \order{0.1}  \left( 1 - \log{ \bigg[ \left( \frac{m_1}{10 \GeV}\right) \left(\frac{\mAp}{3 m_1}\right)^2 \bigg]} \right)
,
\ee
where  we have fixed $x \simeq 20$ to the typical value at freeze-out. For mass-splittings of this size, the exponential Boltzmann factor in Eq.~(\ref{eq:sigmav1}) corresponds to only a mild suppression at temperatures relevant for the freeze-out of $\x_1$. However, at much later times, such as during recombination, the annihilation rate is completely negligible. This significantly alleviates strong bounds derived from distortions of the CMB~\cite{Ade:2015xua,Slatyer:2015jla}. %as a result of ionizing energy injection into the electromagnetic bath~\cite{Ade:2015xua,Slatyer:2015jla}.

%%%
\begin{figure}[t]
\centering
\includegraphics[width=0.5\textwidth]{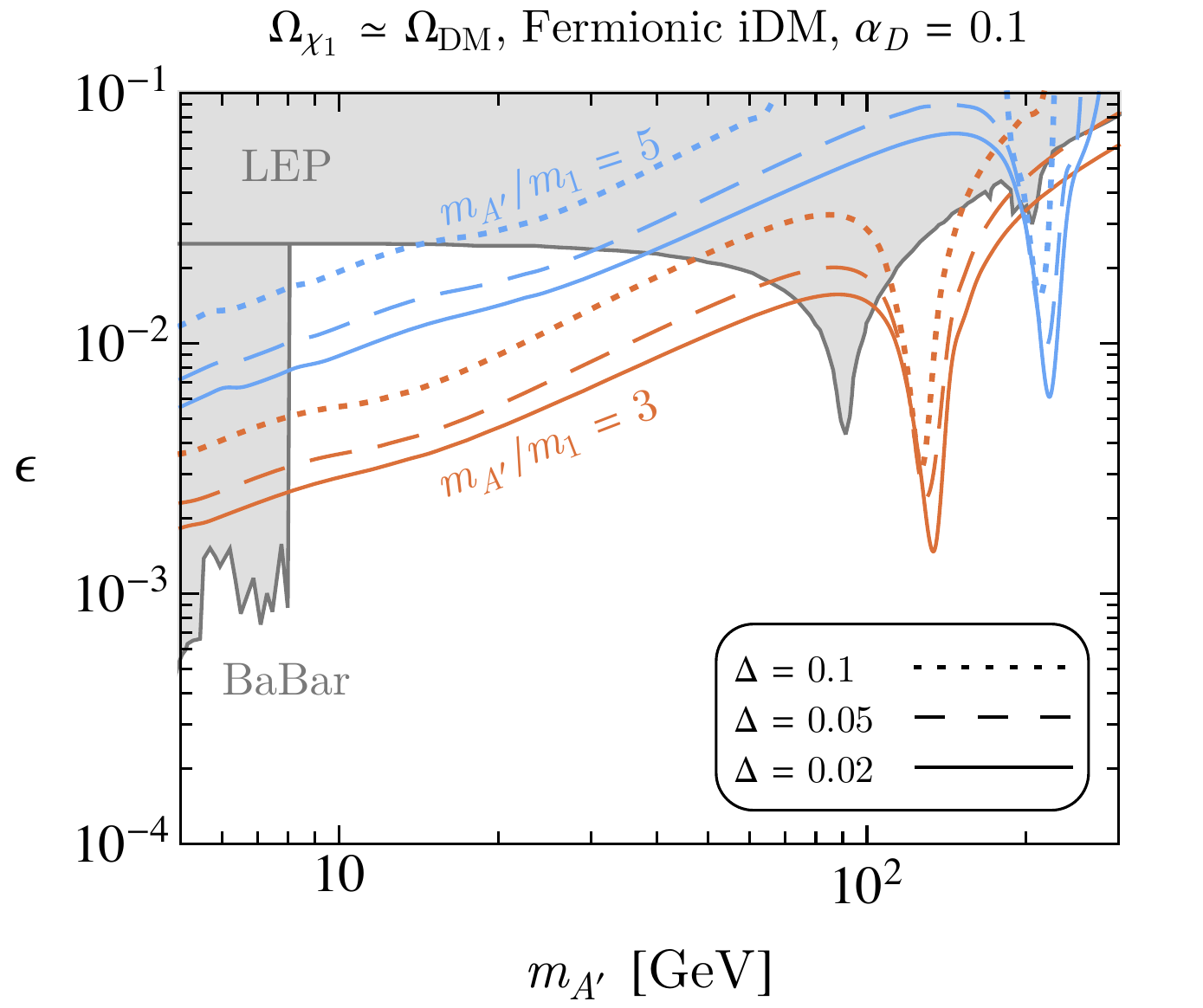}
\caption{Viable parameter space in the $\eps-\mAp$ plane for a model of inelastic dark matter with $\alpha_D = 0.1$. Along the colored contours, $\x_1$ acquires an abundance (through $\x_1 \x_2 \to A^{\prime *} , Z^* \to f \bar{f}$) that is in agreement with the observed dark matter energy density. Solid, dashed, and dotted contours correspond to relative $\x_1-\x_2$ mass-splittings of $\Delta = 0.02$, $0.05$, and $0.1$, respectively, while the orange and blue sets of contours correspond to the $\Ap-\x_1$ mass ratios of $\mAp / m_1 = 3$ and $5$, respectively. Existing constraints from a monophoton search at BaBar~\cite{Aubert:2008as,Lees:2017lec} and precision electroweak measurements at LEP~\cite{Hook:2010tw,Curtin:2014cca} are shown as solid gray regions. The large area of viable and cosmologically motivated parameter space corresponding to $10 \GeV \lesssim \mAp \lesssim 100 \GeV$ and $10^{-3} \lesssim \eps \lesssim \text{few} \times 10^{-2}$ is the primary region of interest in this work.}
\label{fig:IDM_Relic}
\end{figure}
%%%

We numerically solve the Boltzmann equations for the relic abundance of $\x_1$, incorporating hadronic and leptonic final states, as detailed in Refs.~\cite{Griest:1990kh,Gondolo:1990dk,Edsjo:1997bg,Izaguirre:2015zva}. Our results are shown in Fig.~\ref{fig:IDM_Relic}. We do not incorporate intermediate GeV-scale hadronic resonances in this figure since our main focus throughout this work is on $m_1 \gtrsim \text{few} \times \gev$. Although the dark photon predominantly mixes with the SM photon for $\mAp \lesssim \order{10} \GeV$, it has additional non-negligible mixing with the $Z$ for larger masses. A detailed discussion of $\Ap - \gamma , Z$ mixing is given in Appendix~\ref{sec:mixing}. In addition to $\Ap$ exchange, we include processes where the DM annihilates through an intermediate $Z$ ($\x_1 \x_2 \to A^{\prime *} , Z^* \to f \bar{f}$). These contributions have a sizable effect on the DM relic abundance near the $Z$-resonance at $m_1 \simeq m_Z/2$.\footnote{Compared to the relic abundance calculation performed in Refs.~\cite{Izaguirre:2015zva,Izaguirre:2017bqb}, we do not find a significant enhancement of the annihilation rate at $\mAp \simeq m_Z$, due to interference between $\Ap$- and $Z$-exchange diagrams that contribute to to the annihilation amplitude with opposite signs. This is shown explicitly in Eq.~(\ref{eq:interference}) of Appendix~\ref{sec:mixing}.} We have also checked that our analytic results agree well with the numerical output of the publicly available codes FeynRules~\cite{Alloul:2013bka} and MicrOMEGAS~\cite{Belanger:2013oya}. 

In Fig.~\ref{fig:IDM_Relic}, we show regions of parameter space in the $\eps - \mAp$ plane where $\x_1$ freezes out with an abundance that is in agreement with the observed DM energy density for various choices of model parameters. The solid, dashed, and dotted colored contours correspond to the fractional mass-splittings of $\Delta = 0.02$, $0.05$, and $0.1$, respectively, whereas the orange and blue sets of contours correspond to the mass ratios $\mAp / m_1 = 3$ and $\mAp / m_1 = 5$. We have fixed $\alpha_D = 0.1$ throughout. As shown in Fig.~\ref{fig:IDM_Relic}, for fixed $\alpha_D$ and $m_1$, smaller mass-splittings ($\Delta$) and mass ratios ($\mAp / m_1$) enhance the rate for $\x_1 - \x_2$ coannihilations in the early universe and facilitate smaller values of $\eps$ in the cosmologically favored parts of parameter space.\footnote{In tuned regions of parameter space where $\mAp / m_1 \simeq 2$, significant resonant enhancements of the annihilation rate during freeze-out allow for couplings that are smaller by a few orders of magnitude~\cite{Griest:1990kh,Feng:2017drg,Berlin:2018bsc}. We do not consider such scenarios in this study.} Also shown in Fig.~\ref{fig:IDM_Relic} are existing constraints from measurements of electroweak precision observables at LEP~\cite{Hook:2010tw,Curtin:2014cca} and a monophoton search at BaBar~\cite{Lees:2017lec}. These bounds, along with several others, will be discussed in detail in Sec.~\ref{sec:constraints}.

If $\alpha_D$ is comparable to SM gauge couplings, then $\alpha_D \gg \alpha_\text{em} \, \eps^2$ for the cosmologically motivated parameter space of Fig.~\ref{fig:IDM_Relic}. As a result, decays of $\Ap$ to $\x_1 \x_2$ pairs dominate over those directly to SM particles if kinematically accessible. In the limit that $\mAp \gg m_1 \simeq m_2$, the corresponding partial width is
\be
\Gamma (\Ap \to \x_1 \x_2) \simeq \frac{\alpha_D \, \mAp}{3}
~.
\ee
For dark photons produced in terrestrial experiments, such decays are the dominant production mechanism for DM and its excited state. Once a $\x_1 \x_2$ pair is produced in this manner, $\x_1$ leaves the detector as missing energy/momentum. The heavier counterpart, $\x_2$, is unstable and eventually undergoes a three-body decay to DM and a pair of SM particles through an off-shell $\Ap$, i.e., $\x_2 \to \x_1 A^{\prime *} \to \x_1 f \bar{f}$, giving rise to hidden valley-like signatures~\cite{Strassler:2006im}. The rate for $\x_2$ to decay to $\x_1$ and a pair of SM leptons is approximately given by
\be
\label{eq:decaychi2}
\Gamma(\x_2 \to \x_1 \, \ell^+ \ell^- ) \simeq \frac{4\, \eps^2\, \alpha_\text{em} \, \alpha_D \, \Delta^5 m_1^5}{15\pi \, \mAp^4}
~,
\ee
where we have taken $\mAp \gg m_1 \simeq m_2 \gg m_\ell$. For $\eps \ll 1$ and $\Delta \lesssim 0.1$, the proper lifetime of $\x_2$ is easily macroscopic, 
\begin{align}
\label{eq:decaychi2V2}
c \, \tau_{\x_2} &\sim \order{\text{m}} \times \left( \frac{\alpha_D}{0.1} \right)^{-1} \left( \frac{\eps}{10^{-2}} \right)^{-2}  
\nl
&~~
\left( \frac{\mAp}{3 m_1} \right)^4 \left( \frac{\Delta}{0.05} \right)^{-5} \left( \frac{m_1}{10 \GeV} \right)^{-1}
.
\end{align}

In experiments with baselines significantly smaller than the boosted lifetime of $\x_2$ (such as the $B$-factories BaBar and Belle II), the process $\Ap \to \x_1 \x_2$ often registers as an invisible decay. On the other hand, longer baseline experiments can efficiently search for the visible decay products from the long-lived $\x_2$. While previous studies have shown that existing and future $B$-factories and fixed-target experiments will definitively test large regions of the cosmologically motivated parameter space for $\mAp \lesssim 10 \GeV$~\cite{Izaguirre:2017bqb,Berlin:2018pwi}, heavier masses remain relatively unconstrained (see, however, Ref.~\cite{Izaguirre:2015zva}). Amongst the proposed experiments in the search for long-lived and weakly-coupled hidden sectors, those that are able to leverage the high center of mass energy of the LHC (such as the ones discussed in this work) pose a strong advantage in this regard. In the remainder of this work, we focus on the details of $\Ap$ production and the detection of $\x_2$ in the large open region of parameter space corresponding to $10 \GeV \lesssim \mAp \lesssim 100 \GeV$.

%%%%%%%%%%%%%%%%%%%%%%%%%%%%%%%%%%%
%%% EXISTING CONSTRAINTS  
%%%%%%%%%%%%%%%%%%%%%%%%%%%%%%%%%%%

\section{Existing Constraints}
\label{sec:constraints}

For the sake of completeness, we briefly review in this section the multitude of existing constraints on models of light iDM. These constraints can be classified into model-independent bounds and those arising from invisible or visible signals. Below, we will only discuss the most stringent of these that are directly relevant to the models under investigation in this work. For a more complete discussion, we refer the reader to the relevant sections of, e.g., Refs.~\cite{Izaguirre:2017bqb,Berlin:2018pwi}.

Experimental measurements that are independent of unknown dynamics in the hidden sector constitute a highly powerful probe of these models. A prime example of this arises from electroweak precision measurements performed at LEP. The kinetic mixing of Eq.~(\ref{eq:kinmix}) generically leads to various shifts in electroweak precision observables, and recent fits give a model-independent upper bound of $\eps \lesssim 3 \times 10^{-2}$ for $\mAp \lesssim 100 \GeV$~\cite{Hook:2010tw,Curtin:2014cca}. For dark photon masses of $10 \lesssim \mAp \lesssim 100 \GeV$, these are the most stringent existing constraints. Similar tests of electroweak precision observables at the LHC are expected to improved upon these limits by a factor of $\lesssim 2$ in $\eps$~\cite{Curtin:2014cca}.

For $\mAp \lesssim 10 \GeV$, a monophoton search at BaBar places an upper bound on the kinetic mixing of $\eps \lesssim 10^{-3}$. In models of iDM, this signal arises from $e^+ e^- \to \gamma + \Ap \to \gamma + \x_1 \x_2$. The $\x_2$ (along with the $\x_1$) is registered as missing energy provided that it escapes the detector or if the leptons from its decay ($\x_2 \to \x_1 \ell^+ \ell^-$) fall below the required energy thresholds~\cite{Aubert:2008as,Lees:2017lec}. For $\Delta \lesssim 0.1$, we find that $\x_2$ is regarded as an invisible final state in this search for most relevant regions of parameter space. A similar search with much higher luminosity is ultimately expected to be performed at Belle II in the near future, with an improved sensitivity to $\eps$ by as much as an order of magnitude~\cite{Inguglia:2016acz,Ferber:2017zjh,DePietro:2018sgg,Kou:2018nap}. 

Although not the primary focus of this paper, we also briefly discuss existing constraints for sub-GeV hidden sector masses. In particular, for $m_{1,2} \lesssim \mAp \lesssim \text{GeV}$, strong bounds can be derived from searches for energy deposition from $\x_{1,2}$ scatters or $\x_2$ decays in detectors placed downstream of high-intensity beam dump experiments, such as LSND~\cite{Auerbach:2001wg,deNiverville:2011it}, E137~\cite{Bjorken:1988as,Batell:2014mga}, and MiniBooNE~\cite{Aguilar-Arevalo:2017mqx,Aguilar-Arevalo:2018wea}. For sizable $\alpha_D$ and $\x_1-\x_2$ mass-splittings of $\Delta \sim \order{0.1}$, these searches lead to upper bounds of roughly $\eps \lesssim \order{10^{-5}} - \order{10^{-3}}$ for $10 \MeV \lesssim m_1 \lesssim 1 \GeV$, respectively~\cite{Izaguirre:2017bqb,Berlin:2018pwi}. 

%%%%%%%%%%%%%%%%%%%%%%%%%%%%%%%%%%%
%%% IDM RESULTS AT LHC                          
%%%%%%%%%%%%%%%%%%%%%%%%%%%%%%%%%%%

\section{Inelastic Dark Matter at the LHC}
\label{sec:searches}

In this section, we discuss dedicated searches at existing and proposed LHC experiments for visible signals of iDM. We consider on-shell production of dark photons for $\mAp \lesssim \text{few} \times 100 \GeV$. A detailed discussion of the various $\Ap$ production mechanisms will be provided later in \secref{prod}. If $\mAp \gtrsim m_1 + m_2$, once produced, the dark photon promptly decays to a $\x_1 \x_2$ pair. $\x_1$ is cosmologically stable and easily escapes the instrumental geometry without interacting. However, since $\x_2$ is unstable and naturally long-lived, we will focus on dedicated searches for visible displaced vertices arising from its decay, $\x_2 \to \x_1 f \bar{f}$, where $f$ is a SM fermion. We now proceed by outlining proposed experimental searches at ATLAS, CMS, LHCb, MATHUSLA, CODEX-b, and FASER.

\subsection{ ATLAS/CMS}

As is common in many DM models, the production of iDM at the LHC leads to missing (transverse) energy. In the parameter space of interest ($\mAp \lesssim 100 \GeV$, as in Fig.~\ref{fig:IDM_Relic}), the missing energy is typically small ($\met \sim \mAp$). As a result, the detection of such a signal is challenging given the trigger requirements and large background rates. However, if the decay length of the excited iDM state is not too large ($c\tau_{\x_2}\lesssim~\order{\m}$), it often decays inside the ATLAS and CMS detectors leaving an additional visible component to the signal. In the following, we will consider two types of proposed search strategies for these LLPs, looking for either a displaced lepton-jet or a time-delayed signal. 

%%%
\begin{figure*}[t]
\centering
\includegraphics[width=1\textwidth]{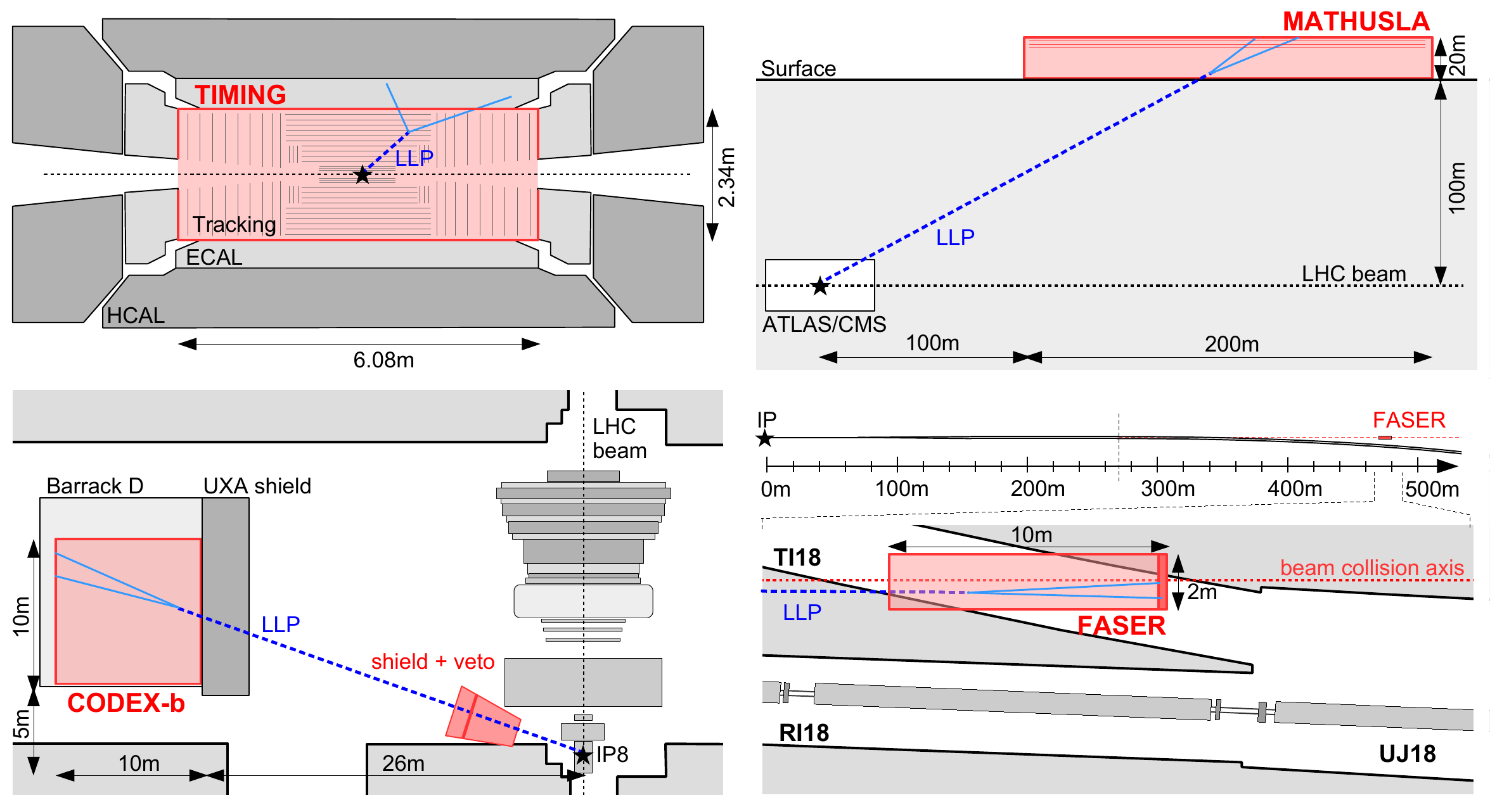} 
\caption{Schematic drawings of a timing layer at CMS (top-left), MATHUSLA (top-right), CODEX-b (bottom-left), and  FASER (bottom-right), along with their locations with respect to the LHC ring. The red shaded region indicates the decay volume for each experiment. 
}
\label{fig:experiments}
\end{figure*}
%%%

\subsubsection{Displaced Muon-Jet}
\label{sec:DMJ}

A displaced decay of the excited iDM state, $\x_2$, inside the LHC detectors is a striking signature and provides a powerful handle for background rejection. Roughly a quarter of the time, $\x_2$ decays into muons when kinematically accessible, leading to a particularly clean signal: a so-called \textit{displaced muon-jet} (DMJ). Such a search has been proposed by the authors of Ref.~\cite{Izaguirre:2015zva}, and in the following we adapt their analysis strategy. 

As suggested in Ref.~\cite{Izaguirre:2015zva}, we utilize the monojet + $\met$ trigger, which requires missing energy balanced by a recoil jet with transverse momentum larger than $p_{T,j}>120 \GeV$. Such a trigger has been used in previous DM searches at both 8~TeV~\cite{Khachatryan:2014rra} and 13~TeV~\cite{Sirunyan:2018ldc} runs of the LHC and we assume that a similar performance can be achieved at a high-luminosity run. 

Due to the small mass-splitting between the iDM states ($\Delta \ll 1$), the $\x_2$ decay products are typically soft, such that $p_{T,\mu} \sim \Delta \times p_{T,\x_2} \ll p_{T,j}$. We require each muon to have a transverse momentum $p_{T,\mu}>5 \GeV$ which roughly corresponds to the threshold for soft muon identification \cite{CMS-DP-2014-020}.  In order to be able to reconstruct the tracks with sufficient precision, they must hit the outer layers of the tracking system. We therefore require a radial displacement of the $\x_2$ decay vertex of $r_{\x_2}<30~\cm$. Finally, we demand that the muon tracks are sufficiently displaced and require a transverse impact parameter of $d_\mu>1~\mm$. In summary, this search strategy for displaced muon-jets at ATLAS and CMS requires

\begin{align}
\textbf{DMJ}: ~~~
& p_{T,j} >120 \GeV
\nl
& p_{T,\mu} > 5 \GeV
\nl
& r_{\x_2} < 30 ~ \cm
\nl
& d_\mu >  1~ \mm
\, .
\end{align}

When estimating the reach, we assume the expected integrated luminosity of the high-luminosity (HL) LHC, $\mathcal{L}=3~\iab$, and that backgrounds can be reduced to a negligible level as argued by the authors of Ref.~\cite{Izaguirre:2015zva}.  

\subsubsection{Time-Delayed Tracks}
 
An alternative search strategy using precision timing has been proposed in Ref.~\cite{Liu:2018wte}. If $\x_2$ decays after traversing a macroscopic distance, its decay products will arrive at the calorimeter delayed in time compared to SM particles that are promptly produced at the interaction point (IP). This time delay is due both to the reduced speed of the massive iDM state, $v_{\x_2}$, and the increased path length of the displaced decay, $l_{\x_2}+l_\ell$ ($\ell$ is a lepton from the decay of $\x_2$), compared to a SM track with path length $l_\text{SM}$. We can estimate the time delay as $\Delta t = l_{\x_2}/v_{\x_2} +l_\ell/c-l_\text{SM}/c$, where  for simplicity we have assumed that the decay products move along straight lines at the speed of light. 

The CMS collaboration recently proposed the installation of a precision timing detector with a resolution of  $\sim 30$ picoseconds. This timing layer would be located in front of the electromagnetic calorimeter (ECAL) with a radial size of $R=1.17~\m$ and extending up to $z=3.04~\m$ along the beam axis. While this upgrade was originally intended for pile-up reduction, its potential implementation in searches for LLPs has been investigated in Ref.~\cite{Liu:2018wte}. A schematic drawing of this setup is shown in the top-left panel of \figref{experiments}.

As in \secref{DMJ}, we require the excited iDM state to decay leptonically. Following Ref.~\cite{Liu:2018wte}, we demand that at least one of the $\x_2$ decay products has a time delay of $\Delta t > 0.3 ~\ns$ and require a recoil jet of $p_{T,j}>30 \GeV$ to timestamp the primary vertex. Since no vertex reconstruction is required for signal identification, this search can make use of the entire decay volume inside the ECAL and access radial and longitudinal displacements of the decay vertex of $r_{\x_2}<1.17~\m$ and  $z_{\x_2}<3.04~\m$, respectively. If there is to be a significant time delay, $\x_2$ should decay sufficiently far away from the primary vertex. We therefore impose a minimal radial decay distance of $r_{\x_2}>5~\cm$. Although the timing layer is expected to have a very low energy threshold, we require the leptons to have a transverse momentum of $p_{T,\ell}>3 \GeV$. This ensures that they travel with light speed along approximately straight trajectories within the magnetic field of the detector. 

We consider two trigger options: a conservative scenario using the conventional trigger and an optimistic  one using a timing-based trigger. For the conservative estimate, we use the conventional monojet + $\met$ trigger and require a recoil jet with $p_{T,j}>120 \GeV$ balancing the missing energy. The cuts in this scenario are
\begin{align}
\label{timing1}
\textbf{Timing ($\met$)} : ~~~
& p_{T,j}  > 120 \GeV
\nl
&p_{T,\ell} >  3 \GeV 
\nl
& \Delta t > 0.3 ~ \ns
 \nl
 & 5~\cm< r_{\x_2} < 1.17 ~ \m 
 \nl
 & z_{\x_2} <  3.04 ~ \m  
 \, .
\end{align}
The presence of a leptonic signature in association with a sizable amount of missing transverse momentum eliminates most potential background sources. While this search does not rely on tracking information for signal reconstruction, it can provide a powerful handle for background rejection. For instance, tracking could be used to veto delayed tracks that are consistent with being produced at or close to the beam line. Alternatively, timing information alone might be sufficient to suppress backgrounds, as discussed in detail by the authors of Ref.~\cite{Liu:2018wte}. In the following, we will assume that SM backgrounds can be reduced to negligible levels. 

The search strategy outlined above is primarily limited by triggering requirements. To explore the full capability of precision timing in LLP searches, we also consider the optimistic case in which the time delay alone can be used for triggering. This allows us to lower the requirement on the transverse momentum of the recoil jet to $p_{T,j}>30 \GeV$, which is sufficient to timestamp the primary vertex. The relevant cuts are then given by
\begin{align}
\textbf{Timing ($\Delta t$)} : ~~~
& p_{T,j} > 30 \GeV
\nl
&p_{T,\ell} >  3 \GeV
\nl
&\Delta t > 0.3 ~ \ns
\nl
& 5~\cm< r_{\x_2} < 1.17 ~ \m
\nl
&z_{\x_2}  <  3.04 ~ \m
\, .
\end{align}
In both cases, we assume an integrated luminosity of $\mathcal{L}=3~\iab$ when estimating the reach.  

\subsection{Displaced Muons at LHCb}

LHCb is a dedicated $B$-physics experiment with coverage in the forward direction and is expected to run with a triggerless readout in the near future, overcoming one of the main challenges of ATLAS and CMS. Furthermore, it has been shown that the detector's vertexing and invariant mass resolution as well as its ability for particle identification makes it an excellent instrument  to search for displaced muon pairs from, e.g., displaced visible decays of dark photons~\cite{Aaij:2017rft,Ilten:2016tkc}. Similar searches are also sensitive to soft energy deposition from the showering and hadronization in confining hidden sectors~\cite{Pierce:2017taw,Buschmann:2015awa}.

We estimate LHCb's sensitivity to displaced muons originating from $\x_2$ decays in models of iDM by applying the baseline post-module search criteria of Ref.~\cite{Ilten:2016tkc},
\begin{align}
\textbf{LHCb} : ~~~
& p_{T,\mu} > 0.5 \GeV
\nl
&p_{\mu} > 10 \GeV
\nl
& 6 ~\mm < r_{\x_2} < 22 ~\mm
\nl
&2 < \eta_{\x_2}  <  5 
\nl
&2 < \eta_{\mu}  <  5 
\, .
\end{align}
The first two requirements on the muon momenta suppress contributions from fake muons, whereas the remaining cuts on the transverse displacement and rapidity ensure that the displaced vertex is sufficiently separated from the beamline and efficiently registered in the Vertex Locator (VELO) with a dimuon identification efficiency of $\sim 50\%$. We note that slightly relaxing the muon momentum thresholds in this analysis could enlarge the discovery potential of LHCb to the signals discussed here. However, lower thresholds also significantly complicate estimates of background contributions to the signal region, and hence, we do not consider this possibility further. 

Ref.~\cite{Ilten:2016tkc} adopted an estimated background of 25 events per mass bin (originating from interactions with the detector material) in a search for resonant dimuons from the visible decays of long-lived dark photons. Although the muons from the three-body decay of $\x_2$ do not reconstruct a resonance, such processes populate a significant fraction of events near the kinematic limit of $m_{\mu \mu} \lesssim \Delta \times m_1$. Strategies to additionally suppress backgrounds by leveraging kinematic handles of such processes at long-baseline experiments has been recently discussed in Ref.~\cite{Ibarra:2018xdl}. A dedicated analysis of this signal and the relevant backgrounds at LHCb is beyond the scope of this work. In estimating the projected sensitivity of LHCb to visible signals of iDM, we will conservatively demand 100 signal events, assuming the integrated luminosity expected by the end of the HL-LHC era, $\mathcal{L}=300~\ifb$. 

\subsection{MATHUSLA}

The sensitivity of LHC experiments that surround the various IPs (such as ATLAS, CMS, and LHCb) is suppressed for small mass-splittings, $\Delta \lesssim \order{0.1}$. In this case, the lifetime of $\x_2$ is sufficiently large that only a small number of decays occur within the inner tracking systems and its decay products are often too soft to overcome limitations from triggering/backgrounds. 

Additional experiments placed further away from the IP that are shielded from large background rates have been proposed to perform dedicated searches for LLPs. The largest of these proposed detectors is MATHUSLA (the MAssive Timing Hodoscope for Ultra-Stable neutraL pArticles)~\cite{Curtin:2018mvb,Alpigiani:2018fgd}. Its planned location is on the surface near ATLAS or CMS. A schematic drawing, showing its position and size, is shown in the top-right panel of \figref{experiments}. The detector consists of a $200~\m \times 200~\m \times 20~\m$ decay volume, positioned $\sim 100~\m$ downstream from the IP and $\sim 100~\m$ above the LHC beam. This large detector volume, covering $\sim 10\%$ of the full solid angle, ensures a sizable geometric acceptance. The rock that is present between the LHC ring and the proposed detector location is expected to reduce QCD backgrounds originating from the IP to a negligible level.  

The decay volume is covered with a scintillating layer to veto incoming  charged particles such as high-energy muons from the IP. Placed on top of the decay volume is a $\sim 5 ~\m$ thick tracking system, which is envisioned to consist of five tracking layers with a timing resolution of $\sim 1~\ns$ and a spatial resolution of $\sim 1~\cm$. This would allow for displaced vertex reconstruction and reliable separation between the upward going LLP signal and the downward going cosmic ray background. Searches for LLPs are assumed to be background free. 

In estimating MATHUSLA's sensitivity, we require $\x_2$ to decay in the decay volume,
\begin{align}
\textbf{MATHUSLA}: ~~~ 
& 100 ~ \m < x_{\x_2} < 120 ~ \m
\nl
 -&100~\m <y_{\x_2} < 100 ~ \m
\nl
 & 100 ~ \m < z_{\x_2} <  300 ~ \m
 \, .
\end{align}
Final state energy thresholds ranging from $200-600 \MeV$ have been discussed in Ref.~\cite{Curtin:2018mvb}. For our baseline analysis, we assume a luminosity of $\mathcal{L}=3~\iab$ and require an energy deposition of $600 \MeV$ per track. In Appendix~\ref{sec:threshold}, we discuss the effect of modifying this threshold on the projected sensitivity.

\subsection{CODEX-b}

While a detector located at the surface necessarily requires a large decay volume to collect a sufficient number of LLP decays, similar physics could be probed with a smaller decay volume if it is located closer to the IP. A possible location for such a detector has been identified in the LHCb cavern. As part of the the upcoming Run 3 upgrade of LHCb, a relocation of the data acquisition system is planned, and a large unoccupied space, shielded by a $3~\m$ thick concrete radiation shield, will become available. CODEX-b (the COmpact Detector for EXotics at LHCb) has been proposed to be constructed in this space~\cite{Gligorov:2017nwh}.  

A schematic drawing of the detector location and size is shown in the bottom-left panel of \figref{experiments}. The decay volume is expected to have dimensions of $10~\m \times 10~\m \times 10~\m$ and to be positioned $\sim 5~\m$ downstream from the IP at a transverse distance of $\sim 26~\m$. This detector design therefore covers approximately $1\%$ of the solid angle. All six sides of the detector would be equipped with a sextet of RPC tracking layers with an effective granularity of $\sim 1~\cm$ and a timing resolution of $\sim 1~\ns$. This would allow for the reconstruction of a displaced vertex inside the decay volume with an $\mathcal{O}(1)$ signal efficiency. We follow Ref.~\cite{Gligorov:2017nwh} and additionally require the track energies to be above a threshold of $\sim 600 \MeV$. In Appendix~\ref{sec:threshold}, we discuss the effect of modifying this threshold on the projected sensitivity.

To reduce background rates to a manageable level, the detector needs to be sufficiently shielded from the large flux of SM particles produced at the IP. In addition to the existing concrete shielding, an additional lead shield located close to the IP and covering the detector geometric acceptance is needed. A $\sim 4.5~\m$ thick lead shield, corresponding to 25 nuclear interaction lengths, would reduce the background rates from most SM particles (e.g., neutrons, kaons, and pions) to negligible levels. The remaining background would predominantly arise from penetrating muons, leading to secondary neutrons or kaons at the far end of the concrete shield which enter the decay volume undetected and produce tracks by scattering on air. An additional scintillating veto to detect charged particles could remove such muon induced events if installed in the lead shield. In the following, we assume that CODEX-b can operate with negligible background. In order to estimate the corresponding sensitivity at the HL-LHC, we take LHCb's ultimate luminosity to be $\mathcal{L}=300~\ifb$. In our analysis, we require that $\x_2$ decays inside the CODEX-b decay volume and that its decay products are sufficiently energetic,
\begin{align}
\textbf{CODEX-b}: ~~~
&26 ~ \m < x_{\x_2} <  36 ~ \m
\nl
-&3 ~ \m < y_{\x_2} <  7 ~ \m
\nl
& 5 ~ \m < z_{\x_2} <  15 ~ \m
\nl
& E_\text{track} > 600 \MeV
\, .
\end{align}

\subsection{FASER}
\label{sec:faser}

New physics at high-energy colliders, such as the LHC, has traditionally been expected in the high transverse momentum region. This is also the case for the experiments discussed above, which are sensitive to LLPs that are produced with large transverse momentum. However, if new particles are light and weakly-coupled, this focus may be misguided. In this case, large event rates are only available at low transverse momentum that is  comparable to the mass of the LLP, $p_T \sim M_\text{LLP}$. Such particles are predominantly produced in the very forward direction, collimated around the beam collision axis, and hence escape detection in typical LHC detectors. Moreover, due to their weak coupling to the SM and the large boost expected in the forward direction, these particles are naturally long-lived and travel a macroscopic distance before decaying. A detector placed in the very forward region along the beam collision axis may therefore be optimal to detect these decays. FASER (the ForwArd Search ExpeRiment) is an experiment designed to take advantage of this opportunity and search for light LLPs in the very forward region. Previous studies have established FASER's potential to discover light new particles~\cite{Feng:2017uoz,Feng:2017vli,Kling:2018wct,Feng:2018noy}

As shown in the bottom-right panel of \figref{experiments}, FASER would be placed along the beam collision axis, several hundred meters downstream of the ATLAS or CMS IP (and after the LHC tunnel begins to curve). A particularly promising location has been identified a few meters outside of the main LHC tunnel, $480~\m$ downstream from the ATLAS IP, in the side tunnel TI18. This space was formerly used to connect the SPS and LEP tunnels but is currently empty and unused. At this location, the beam collision axis intersects with TI18 close to where it merges with the main LHC tunnel at the construction hall UJ18. For concreteness, we assume that the decay volume of FASER has a cylindrical shape with a depth of $D=10~\m$ and a radius of $R=1~\m$. 

Long-lived particles that are produced in the very forward direction and decay in the detector typically have very large energies on the order of $\sim\tev$. The energetic decay products lead to a striking signature at FASER consisting of charged tracks with very high energy, originating from a vertex inside the detector and with a combined momentum pointing back to the IP. A detector that aims to make use of kinematic features to distinguish signal from background therefore needs to be able to measure the individual tracks with sufficient resolution and identify their charges. A tracking-based technology, supplemented with a magnet and calorimeter to allow for energy measurements, would make up the key components of the FASER detector.

The shielding provided by the rock that surrounds the detector's location as well as the forward LHC infrastructure consisting of magnets and absorbers would eliminate most potential background processes. The only known particles that can transport TeV energies through $\sim 100~\m$ of rock between the IP and FASER are muons and neutrinos. A detailed analysis using FLUKA, taking into account the exact layout of the LHC tunnels and models of radiation-matter interactions, has shown that the dominant source of background is radiative processes associated with muons from the $pp$ collision debris~\cite{FLUKAstudy}. Such backgrounds can be identified by the presence of a high-energy muon traversing the full detector and can be suppressed by using a scintillating charged particle veto layer at the front of the detector. Additional backgrounds from neutrino interactions with the detector are small and generally have different kinematics. The study has also shown that no high-energy particles are expected to enter FASER from infrastructure-induced backgrounds, such as proton showers in the dispersion suppressor or from beam-gas interactions. In the following, we assume that backgrounds can be reduced to negligible levels.

%%%
\begin{figure}[t]
\centering
\includegraphics[width=0.48\textwidth]{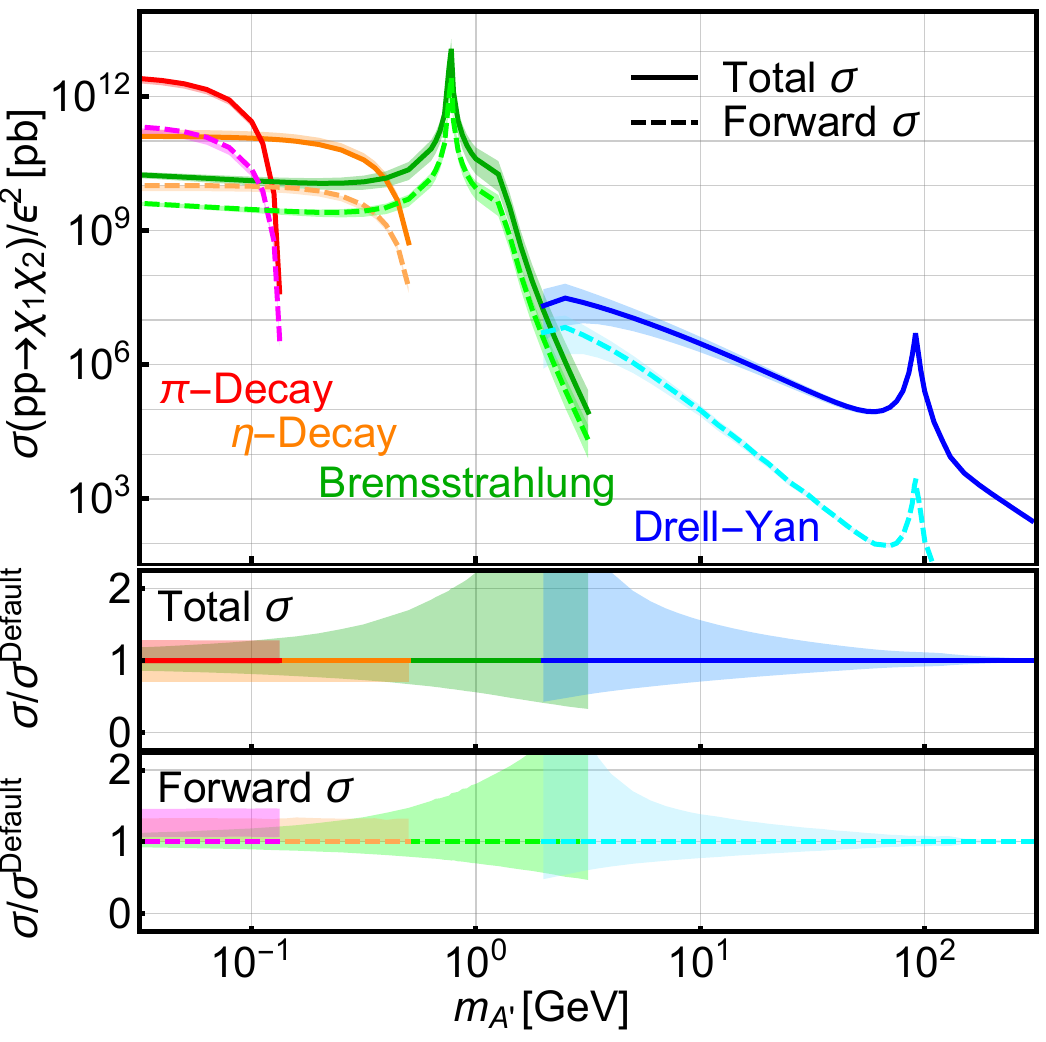}
\caption{Top panel: Inelastic dark matter production cross section, $\sigma(pp \to \x_1 \x_2)$, per $\eps^2$ as a function of the dark photon mass, $\mAp$. We show the total cross section (solid) and the very forward cross section within an angle $\theta_{\x_2} \simeq p_T/p_z<2~\mrad$ of the beam collision axis (dashed). The corresponding systematic uncertainties are shown as the shaded regions. The corresponding relative uncertainties with respect to the central prediction for the total and forward cross section are shown in the central and bottom panels. Here, we fix the model parameters to $\mAp / m_1 = 3$, $\Delta=0.1$, $\alpha_D=0.1$, and $\eps=10^{-3}$. Note that the cross section per $\eps^2$ only depends mildly on these choices.}
\label{fig:crosssection}
\end{figure}
%%%

In order to reduce the trigger rate at low energies, FASER requires a large visible energy deposition from LLP decay products, i.e., $E_\text{vis} \gtrsim 100 \GeV$. For many models, this choice is dictated by the kinematics of the signal, while for iDM, lower energy thresholds are optimal for small DM mass-splittings. In Appendix~\ref{sec:threshold}, we discuss the effect of modifying this threshold on the projected sensitivity. 

In estimating the projected sensitivity of FASER, we assume a luminosity of $\mathcal{L}=3~\iab$. We also require that the  excited iDM state, $\x_2$, decays within the detector volume and that its decay products are sufficiently energetic,
\begin{align}
\textbf{FASER}: ~~~    
& r_{\x_2} <  1~\m  
\nl
& 470 ~ \m < z_{\x_2} <  480 ~ \m
\nl
& E_\text{vis} > 100 \GeV
\, .
\end{align}
%

%%%%%%%%%%%%%%%%%%%%%%%%%%%%%%%%%%%
%%% IDM PRODUCTION                            
%%%%%%%%%%%%%%%%%%%%%%%%%%%%%%%%%%%

\section{Production at the LHC} 
\label{sec:prod}

At the LHC, the iDM states, $\x_1$ and $\x_2$, are pair-produced from the prompt decays of dark photons. In the top panel of \figref{crosssection}, we present the $\x_1 \x_2$ production cross section per $\eps^2$, $\sigma(pp \to \x_1 \x_2)/\eps^2$, as a function of the dark photon mass, $\mAp$, for the dominant production channels: meson decays, dark Bremsstrahlung, and Drell-Yan. The solid lines correspond to the total production cross section in the forward hemisphere ($p_{z, \Ap} > 0$). The lighter dashed lines instead show the production cross section in the very forward region, requiring $\x_2$ (from the decay $\Ap \to \x_1 \x_2$) to be within $2~\mrad$ of the beam collision axis,  which corresponds to the angular acceptance of FASER. The systematic uncertainties of the various production rates are shown as shaded regions around the central prediction. The central and bottom panels show the corresponding relative uncertainties for total and forward production, respectively. 

If the dark photon is lighter than a few hundred MeV, it can be produced from the on-shell decays of pseudoscalar mesons, such as neutral pions and $\eta$-mesons. Following Ref.~\cite{Feng:2017uoz}, we generate the meson spectra using the Monte Carlo code EPOS-LHC~\cite{Pierog:2013ria}, as implemented in the simulation package CRMC~\cite{CRMC}, and subsequently decay the mesons into dark photons, i.e., $\pi^0,\eta \to \gamma \Ap$. The systematic uncertainties are estimated by comparing to the generators QGSJET-II-04~\cite{Ostapchenko:2010vb} and SIBYLL~2.3~\cite{Riehn:2015oba} and are at the $20\%$ level for both the total and forward cross section.

Heavier dark photons above the meson mass threshold are mainly produced via dark Bremsstrahlung and Drell-Yan. We model forward dark Bremsstrahlung using the  Fermi-Weizs{\"a}cker-Williams approximation~\cite{Blumlein:2013cua}, as outlined in Ref.~\cite{Feng:2017uoz}. In particular, we follow Refs.~\cite{deNiverville:2016rqh,Faessler:2009tn} by using a proton form-factor which incorporates off-shell mixing with vector mesons, such as $\rho$ and $\omega$ mesons, leading to an enhanced production rate around $\mAp \simeq 775 \MeV$. The validity requirements of the Fermi-Weizs{\"a}cker-Williams approximation ($E_{\Ap} , E_\text{beam}-E_{\Ap} \gg m_{p} , \mAp , p_{T,\Ap}$) are naturally fulfilled for highly energetic and forward dark photons within the FASER acceptance. However, the approximate formalism is no longer valid for larger $p_{T , \Ap}$ ($\gg \Lambda_{QCD}$) and we therefore  do not include Bremsstrahlung production for CODEX-b or MATHUSLA. Theoretical uncertainties in the Bremsstrahlung calculation are estimated by varying the cut on the dark photon transverse momentum, $p_{T,\Ap}<1 \GeV$, by a factor of two. Note that roughly $10\%$ of all dark photons from meson decays and Bremsstrahlung are produced in the very forward direction (corresponding to the geometric acceptance of FASER), as shown by the dashed lines in \figref{crosssection}.

%%%
\begin{figure*}[t]
\centering
\includegraphics[width=0.49\textwidth]{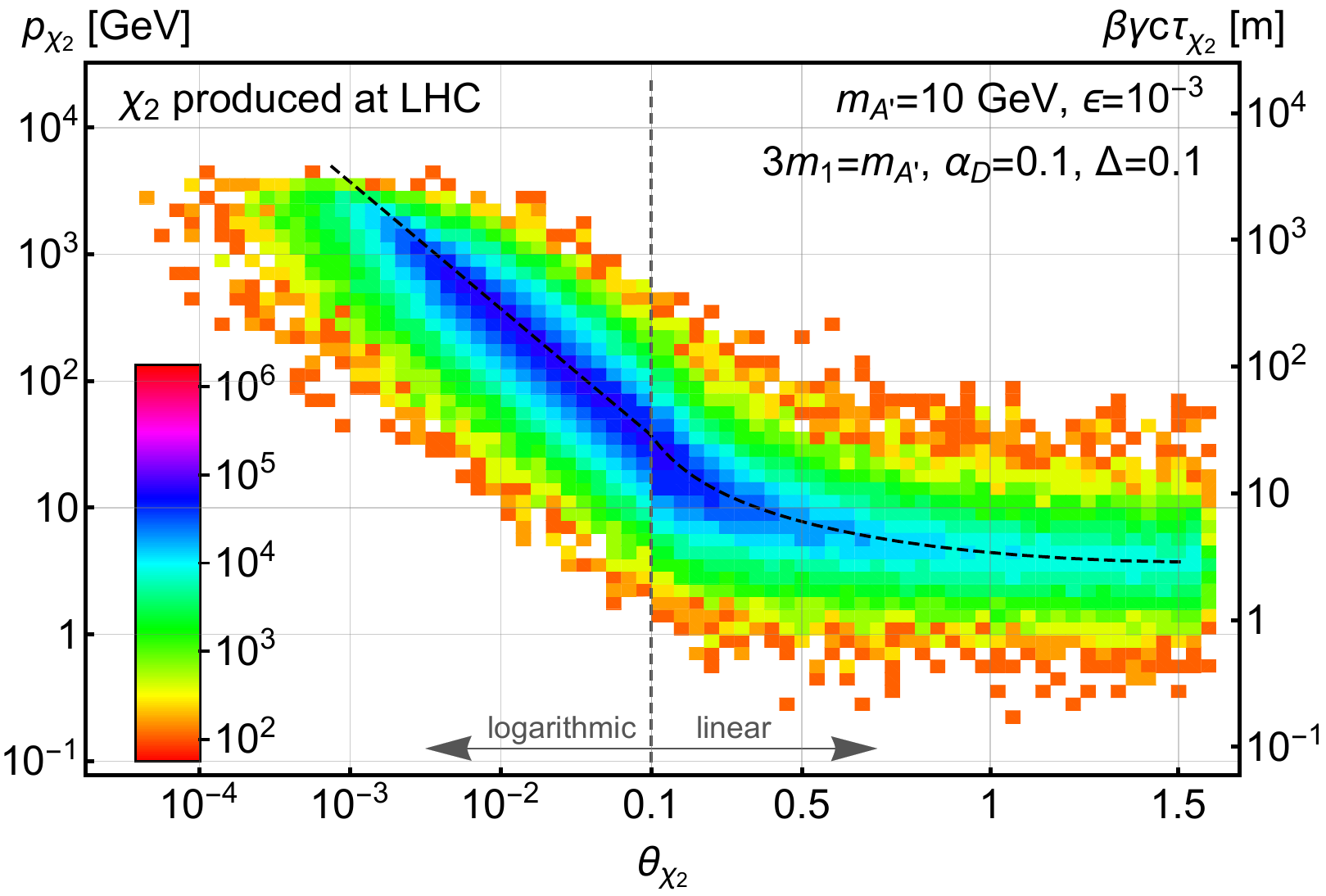}
\includegraphics[width=0.49\textwidth]{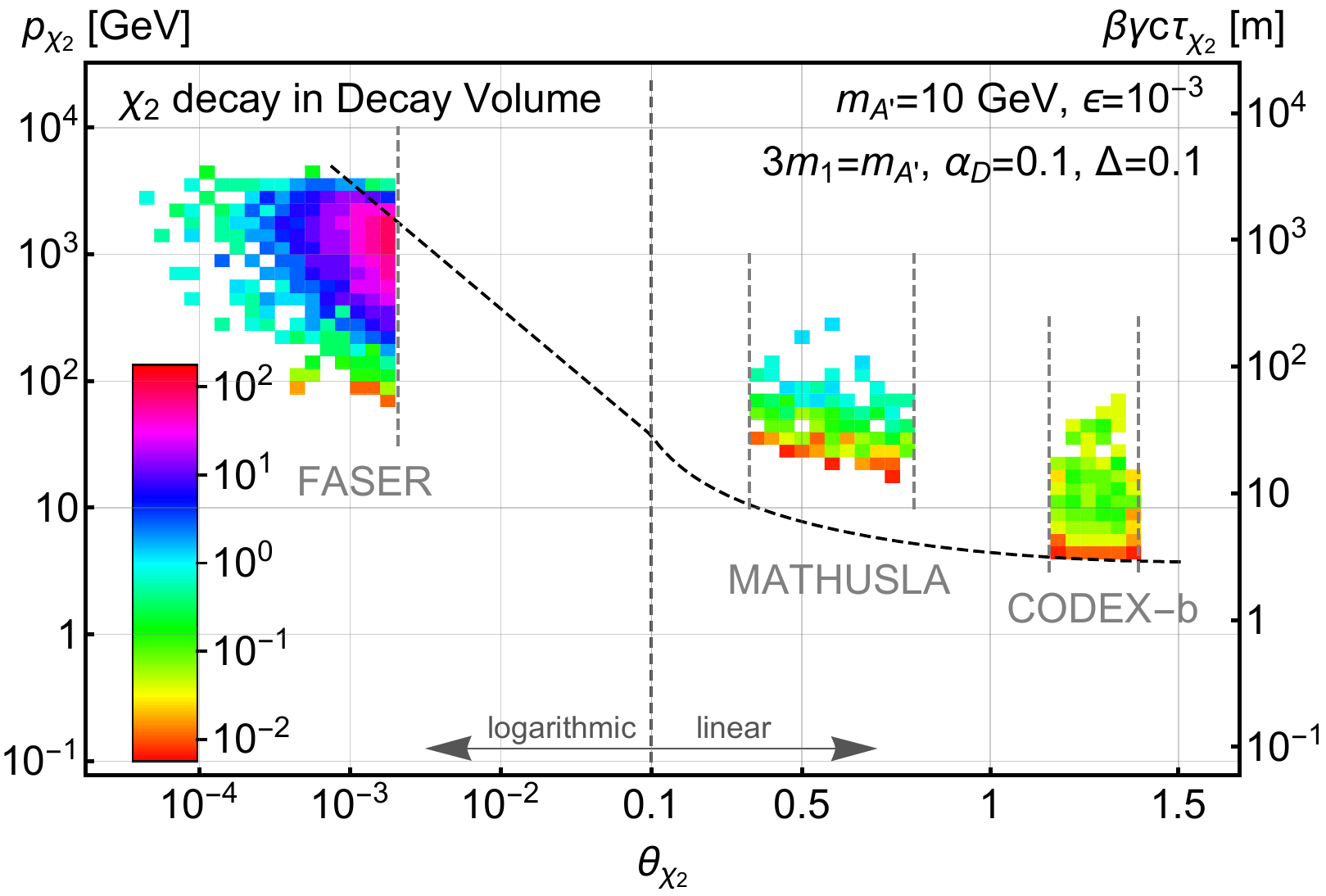}
\caption{
In the $p_{\x_2} - \theta_{\x_2}$ plane, the kinematic distribution of the excited iDM state, $\x_2$, produced at the 13 TeV LHC (left) and the subset of these events where $\x_2$ decays within the detector volume of FASER, MATHUSLA, and CODEX-b (right).  $p_{\x_2}$ and $\theta_{\x_2}$ are the momentum of $\x_2$ and its angle with respect to the beam axis, respectively. Both panels show the number of particles produced in one hemisphere ($p_{z, \x_2} > 0$) with an integrated luminosity of $3~\iab$ for FASER and MATHUSLA and $300 ~ \ifb$ for CODEX-b (the luminosity is fixed only to $3~\iab$ in the left panel). We have fixed $\mAp=10 \GeV$, $m_1 = \mAp / 3$, $\Delta=0.1$, $\eps=10^{-3}$, and $\alpha_D=0.1$. The right-vertical axis indicates the boosted decay length of $\x_2$. The black dashed lines correspond to $p_{T,\x_2} = \frac{1}{2}  \sqrt{\mAp^2-4m_1^2}$, and the gray dashed vertical lines in the right panel show the angular coverage of the experiments. 
}
\label{fig:PvsT}
\end{figure*}
%%%

We follow a different approach to estimate the dark photon production rate at higher transverse momentum, which is relevant for MATHUSLA and CODEX-b. According to the model of vector meson dominance, the photon couples to hadronic states through mixing with intermediate vector mesons. Such interactions also induce an effective mixing between the dark photon and vector mesons, such as the $\rho$. As a result, any physical process producing vector mesons should also lead to the production of dark photons. We perform a rough estimate of the resulting dark photon spectrum by rescaling the inclusive $\rho$ meson spectrum provided by the Monte Carlo generator EPOS-LHC by the appropriate mixing factors. We discuss this in more detail in \appref{VectorMesonMixing}. 

At larger momentum transfers, Drell-Yan processes, such as $q\bar{q} \to \Ap, Z \to \x_1 \x_2$,  also contribute to iDM production. The results were obtained using MadGraph~5~\cite{Alwall:2014hca} and PYTHIA~8~\cite{Sjostrand:2014zea} using the parton distribution function (PDF) NNPDF~3.1~NNLO~\cite{Ball:2017nwa} and the FeynRules model file of Ref.~\cite{Curtin:2014cca}. This model file includes the general $\Ap - \gamma , Z$ mixing angles and masses, as discussed in Appendix~\ref{sec:mixing}, as well as interference effects between the $\Ap$ and $Z$ contributions. As seen in \figref{crosssection}, $\Ap-Z$ mixing leads to an enhancement of the production cross section for $\mAp \simeq m_Z$. Interference effects are typically small except around the $Z$-peak when the two resonances overlap, $\mAp \pm \Gamma_{\Ap} \simeq m_Z \pm \Gamma_Z$. In this case, the cross section also depends on the value of $\alpha_D$ which determines the width of the dark photon and hence the size of the interference. The overall cross section in \figref{crosssection} approximately scales with the kinetic mixing parameter as $\sigma \propto \eps^2$. Small deviations from this simple scaling occur if $\mAp\simeq m_Z$ and $\eps$ is sufficiently large such that $\Gamma(Z \to \x_1 \x_2) \propto \eps^2 \alpha_D$ significantly alters the total $Z$ decay width. However, $\eps$ values of this size are already excluded by LEP~\cite{Hook:2010tw,Curtin:2014cca}. In modeling Drell-Yan production, the dominant systematic uncertainty arises from the choice of scale at which the PDFs are evaluated. We estimate the corresponding scale uncertainty by varying the renormalization and factorization scales ($\mu$) by a factor of two around their central values, $\mu^2 = \mAp^2$. As shown by the blue shaded region in the central and bottom panels of \figref{crosssection}, this theoretical uncertainty is larger for small dark photon masses and becomes $\mathcal{O}(1)$ for $\mAp \simeq 2 \GeV$.

While GeV-scale dark photons (and the $\x_1 \x_2$ pairs from their decays) that are produced from exotic meson decays or Bremsstrahlung are typically very collimated around the beam collision axis, heavier states have a broader spectrum. This is illustrated in the left panel of \figref{PvsT} where we show the $\x_2$ kinematic distribution from Drell-Yan production in the $p_{\x_2}- \theta_{\x_2}$ plane for $\mAp=10 \GeV$, $ \mAp / m_1 = 3$, $\Delta = 0.1$, $\eps = 10^{-3}$, and $\alpha_D = 0.1$. This representative set of parameters can be probed by CODEX-b, FASER, and MATHUSLA, as shown by the circled gray star in the top-left panel of Fig.~\ref{fig:IDM1}. The transverse momentum of $\x_2$ originates mainly from the decay of the dark photon and therefore peaks around $p_{T, \x_2} \sim(\mAp^2-4m_1^2 )^{1/2} / 2$. Larger masses lead to a more broad and energetic  $p_T$ spectrum. The right-vertical axis of  \figref{PvsT} indicates the boosted decay length of $\x_2$. 

%%%
\begin{figure*}[t]
\centering
\includegraphics[width=0.48\textwidth]{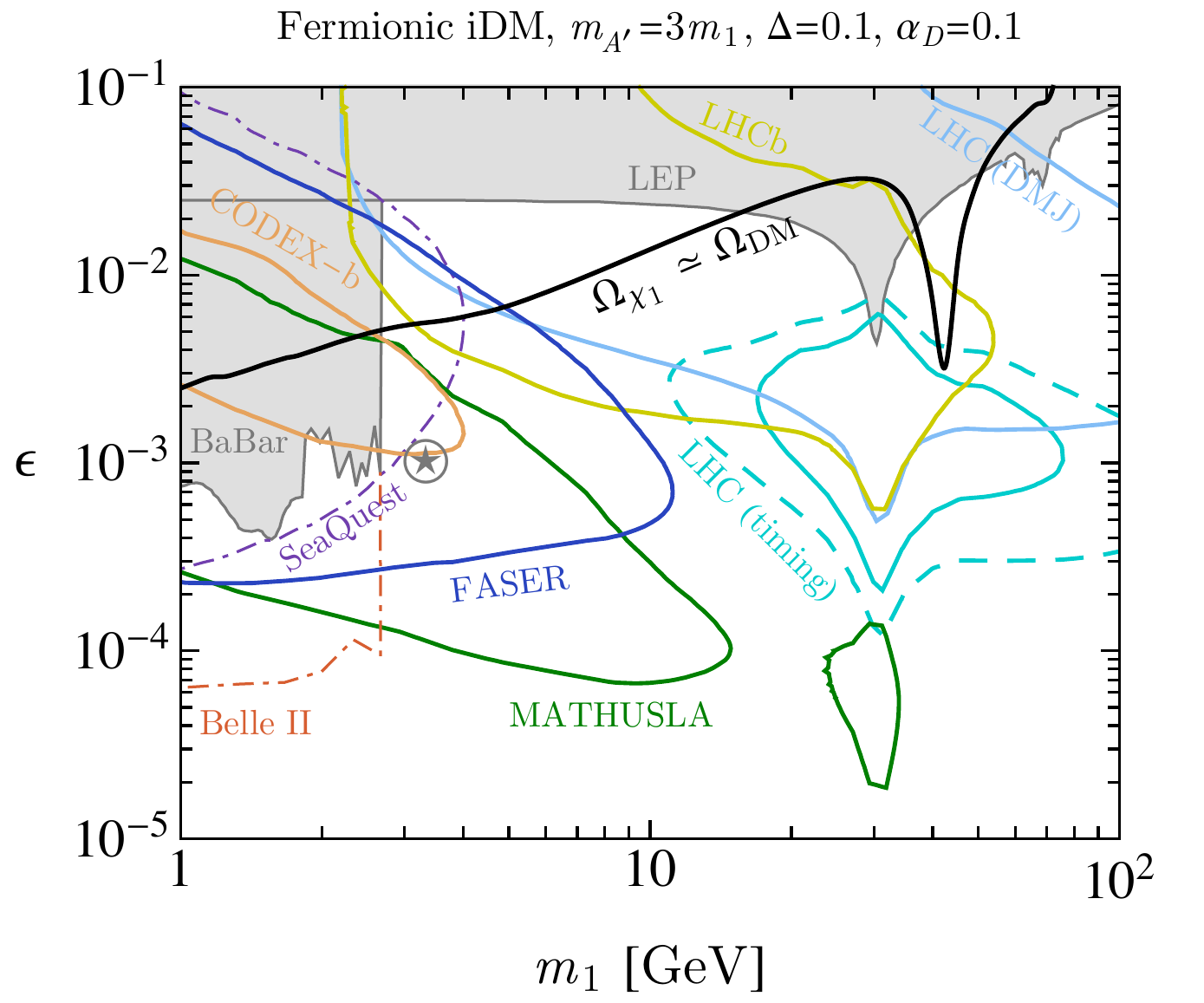}
\includegraphics[width=0.48\textwidth]{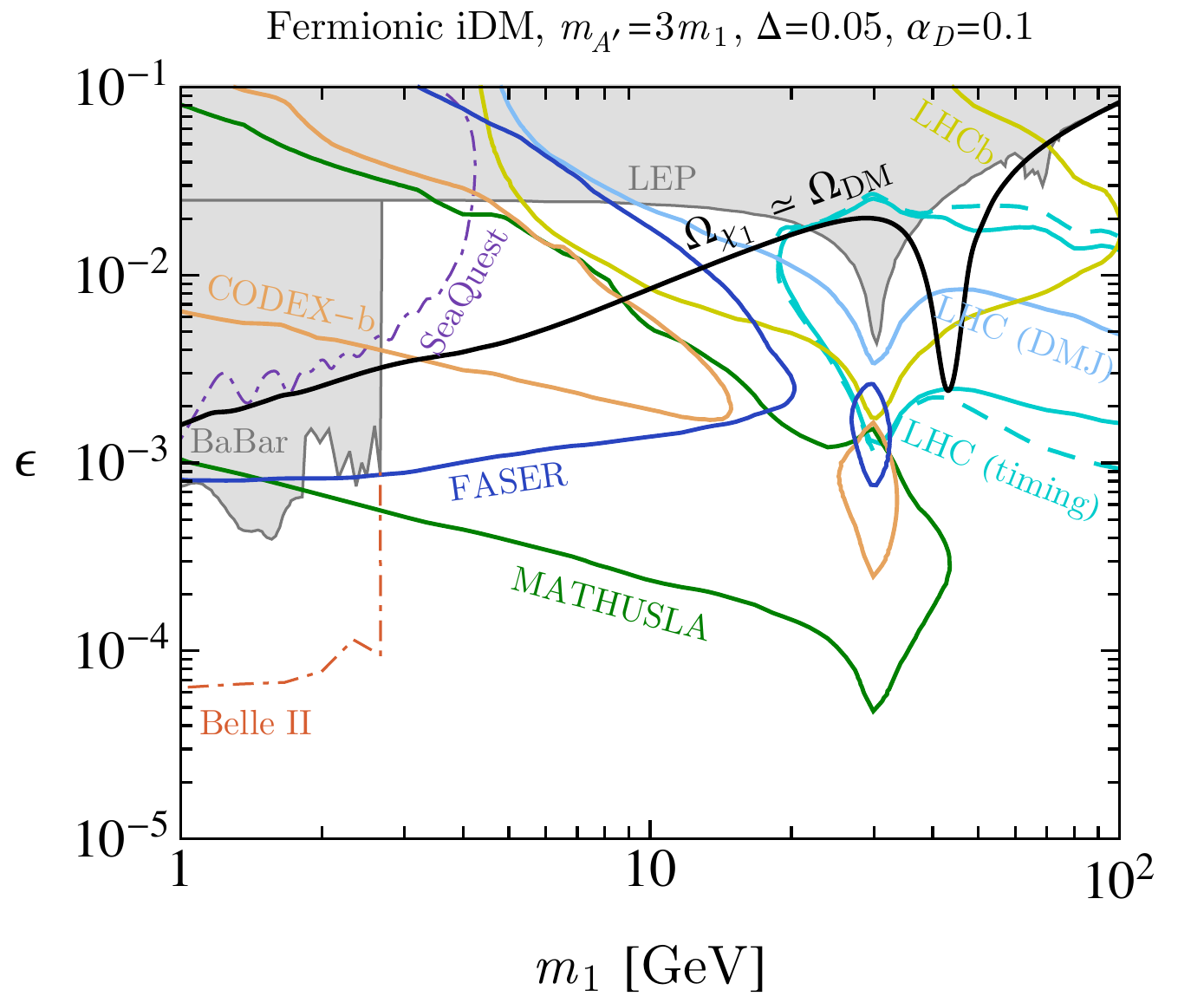}
\includegraphics[width=0.48\textwidth]{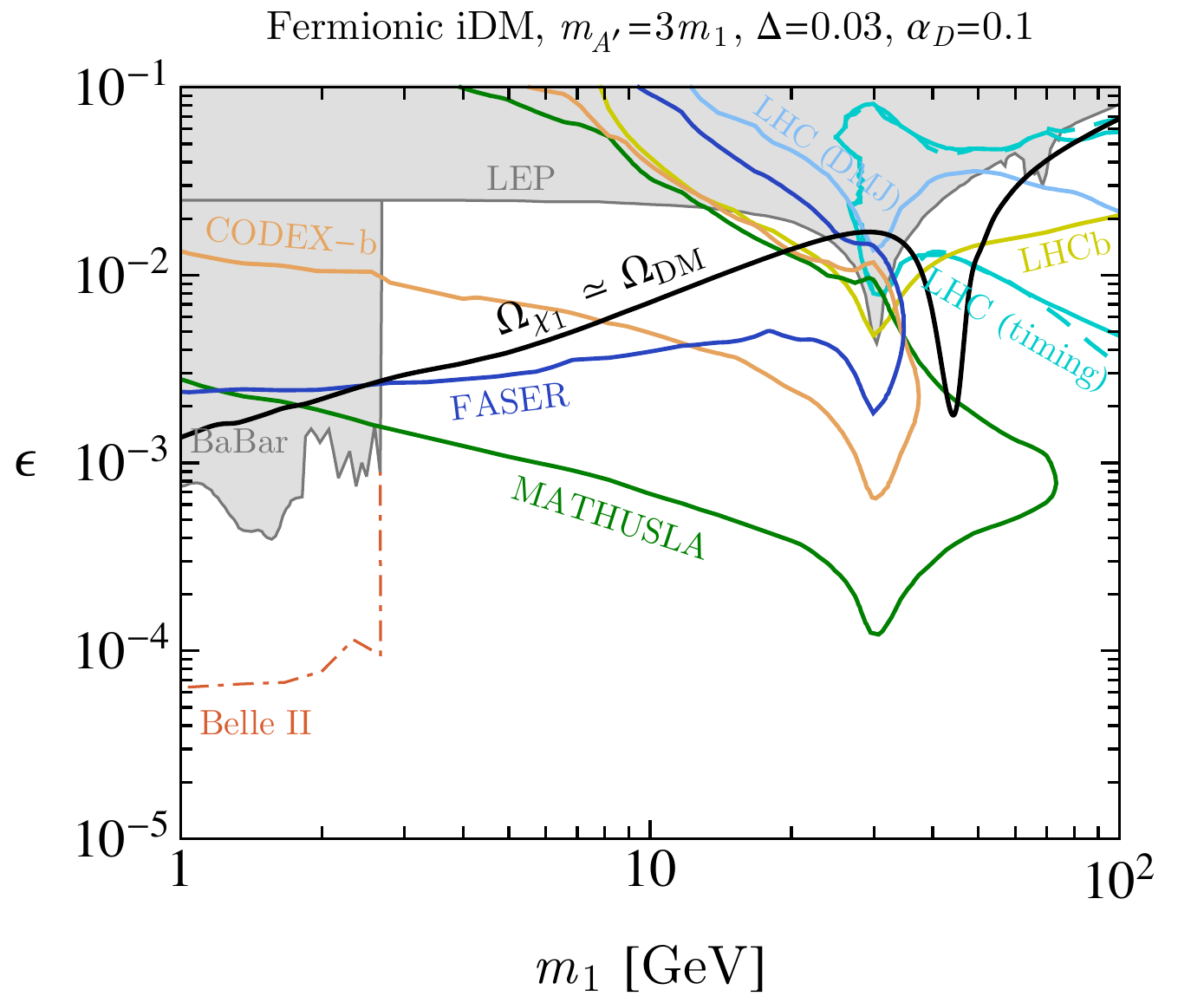}
\includegraphics[width=0.48\textwidth]{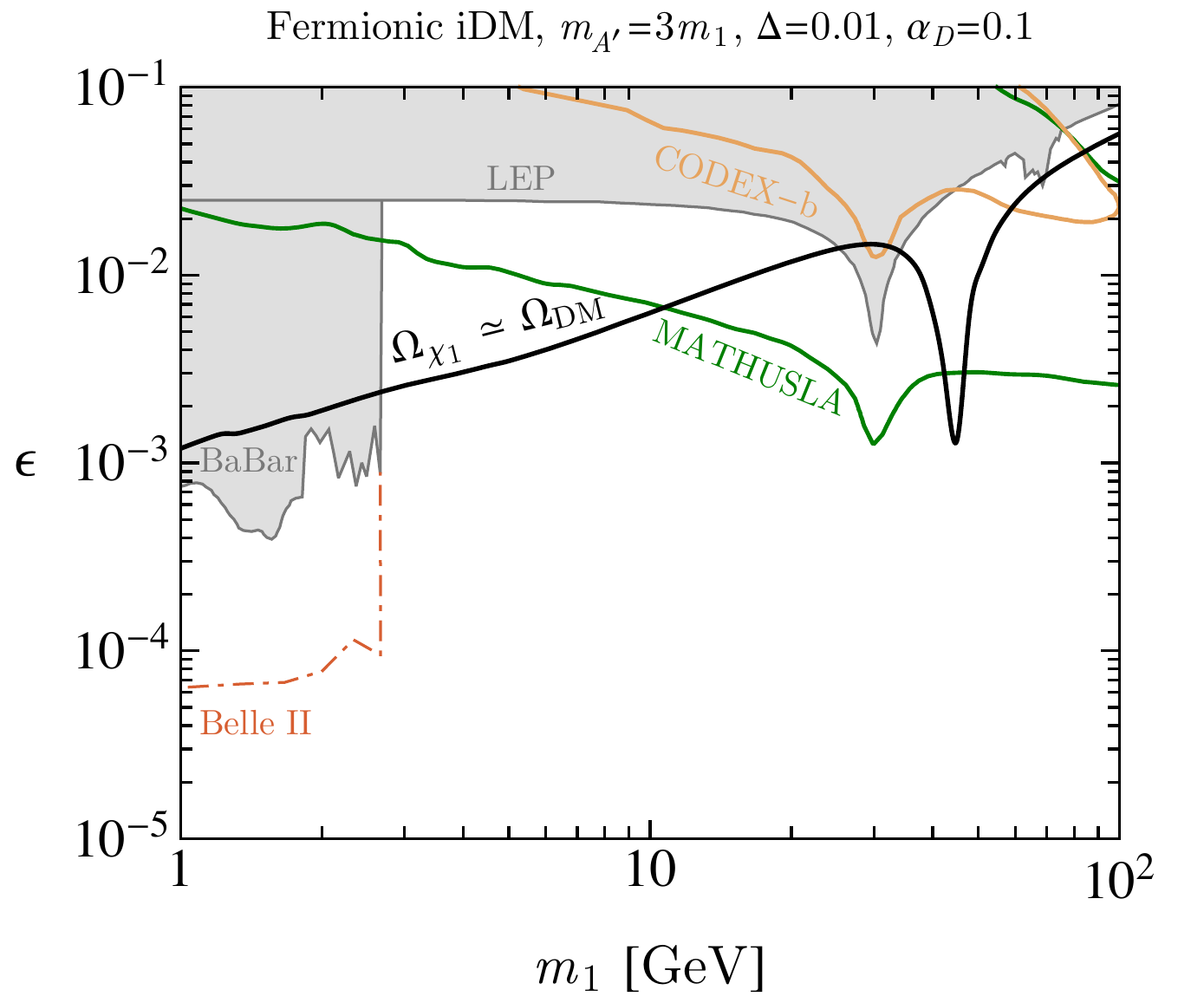}
\caption{Existing constraints (shaded gray) and projected sensitivities (color) to models of fermionic inelastic dark matter in the $\eps-m_1$ plane for $\mAp / m_1 = 3$, $\alpha_D = 0.1$, and various choices of the $\x_1 - \x_2$ fractional mass-splitting ($\Delta = 0.1$, 0.05, 0.03, and 0.01). Along the black contour, the abundance of $\x_1$ matches the observed dark matter energy density. The solid gray regions are excluded from LEP~\cite{Hook:2010tw,Curtin:2014cca} and BaBar~\cite{Lees:2017lec}. The solid colored contours show the projected reach of various proposed searches for displaced vertices at the LHC, such as at ATLAS and CMS (light blue)~\cite{Izaguirre:2015zva}, LHCb (yellow)~\cite{Aaij:2017rft,Ilten:2016tkc,Pierce:2017taw}, CODEX-b (orange)~\cite{Gligorov:2017nwh}, FASER (dark blue)~\cite{Feng:2017uoz}, and MATHUSLA (green)~\cite{Chou:2016lxi}. We also show the sensitivity of a precision timing search at CMS~\cite{Liu:2018wte}, utilizing a conventional monojet (solid cyan) or an optimistic timing-based (dashed cyan) trigger. See the corresponding subsections of Sec.~\ref{sec:searches} for further details. Also shown in dot-dashed are the projected sensitivities of Belle II (red)~\cite{Kou:2018nap} and SeaQuest (purple)~\cite{Berlin:2018pwi}. In the top-left panel, the circled gray star corresponds to the choice of model parameters in Fig.~\ref{fig:PvsT}. 
}
\label{fig:IDM1}
\end{figure*}
%%%

%%%
\begin{figure*}[t]
\centering
\includegraphics[width=0.48\textwidth]{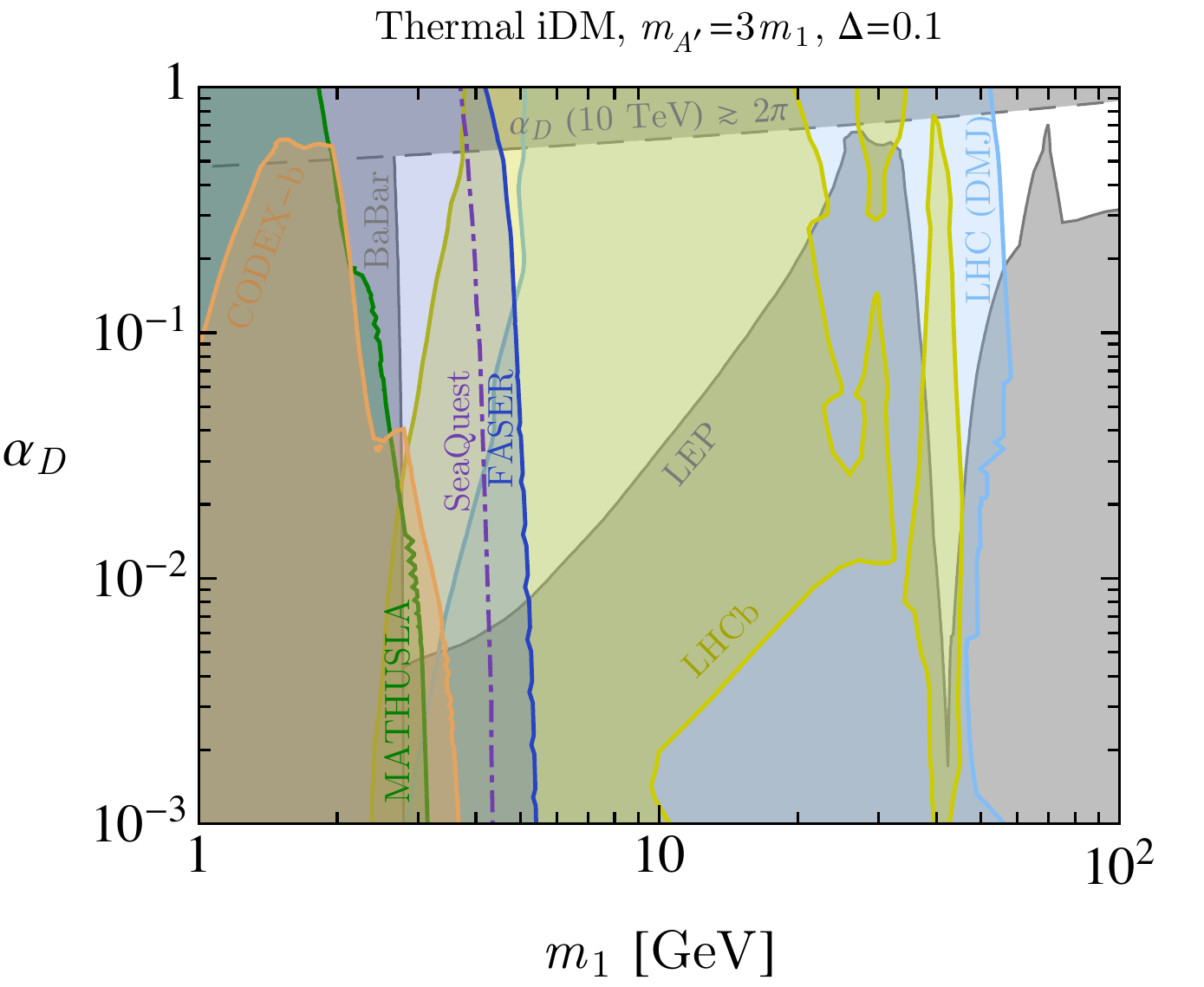}
\includegraphics[width=0.48\textwidth]{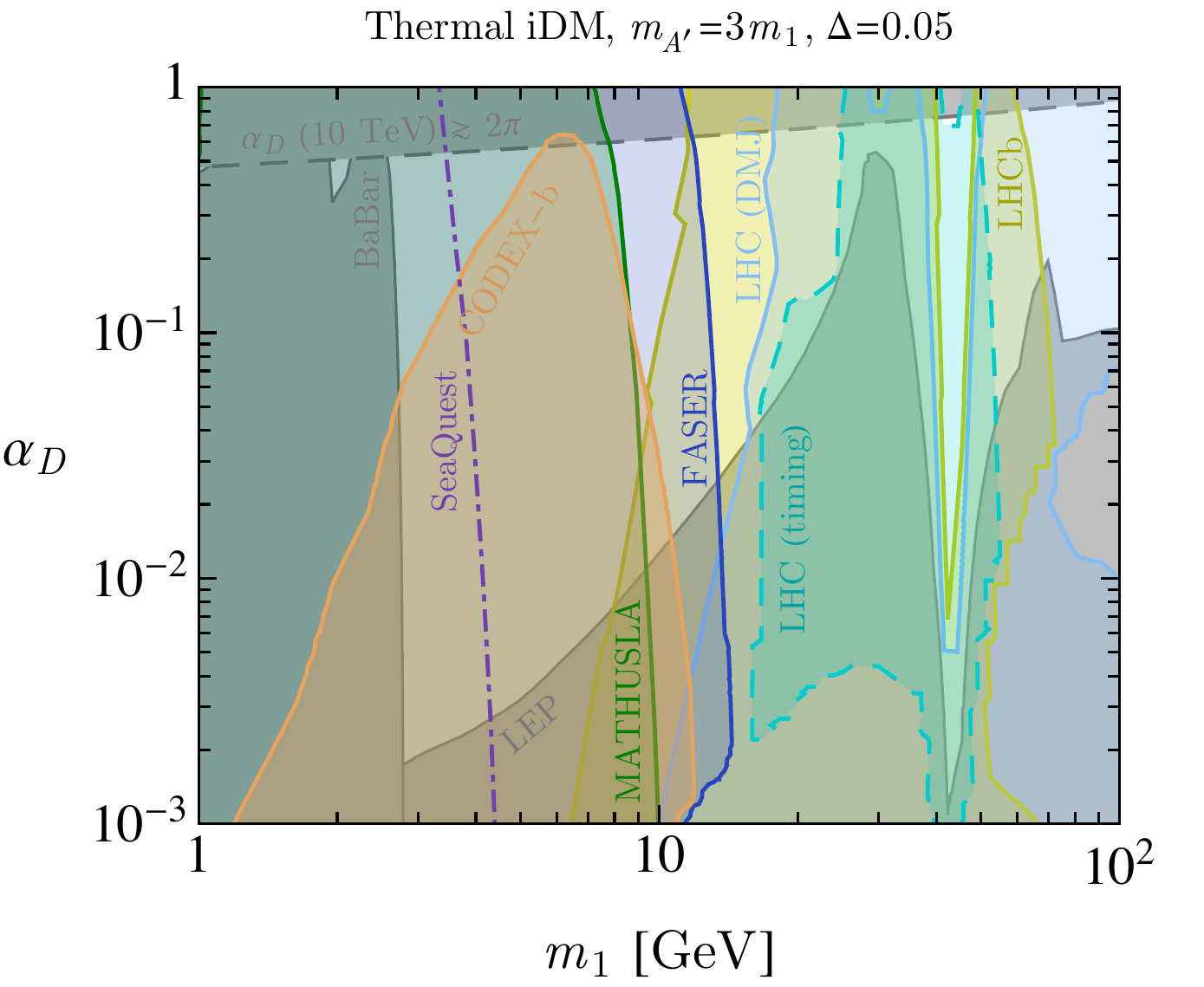}
\includegraphics[width=0.48\textwidth]{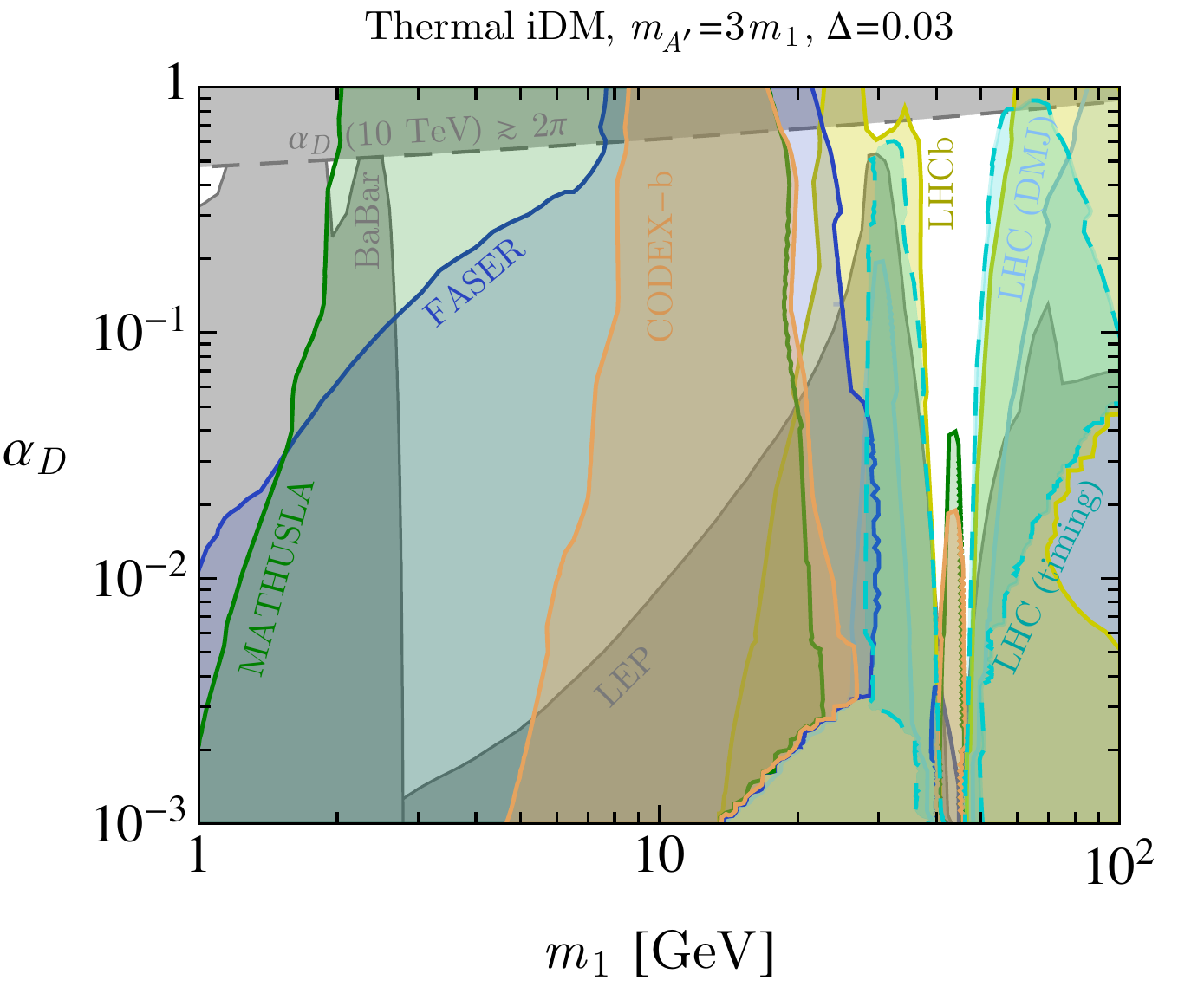}
\includegraphics[width=0.48\textwidth]{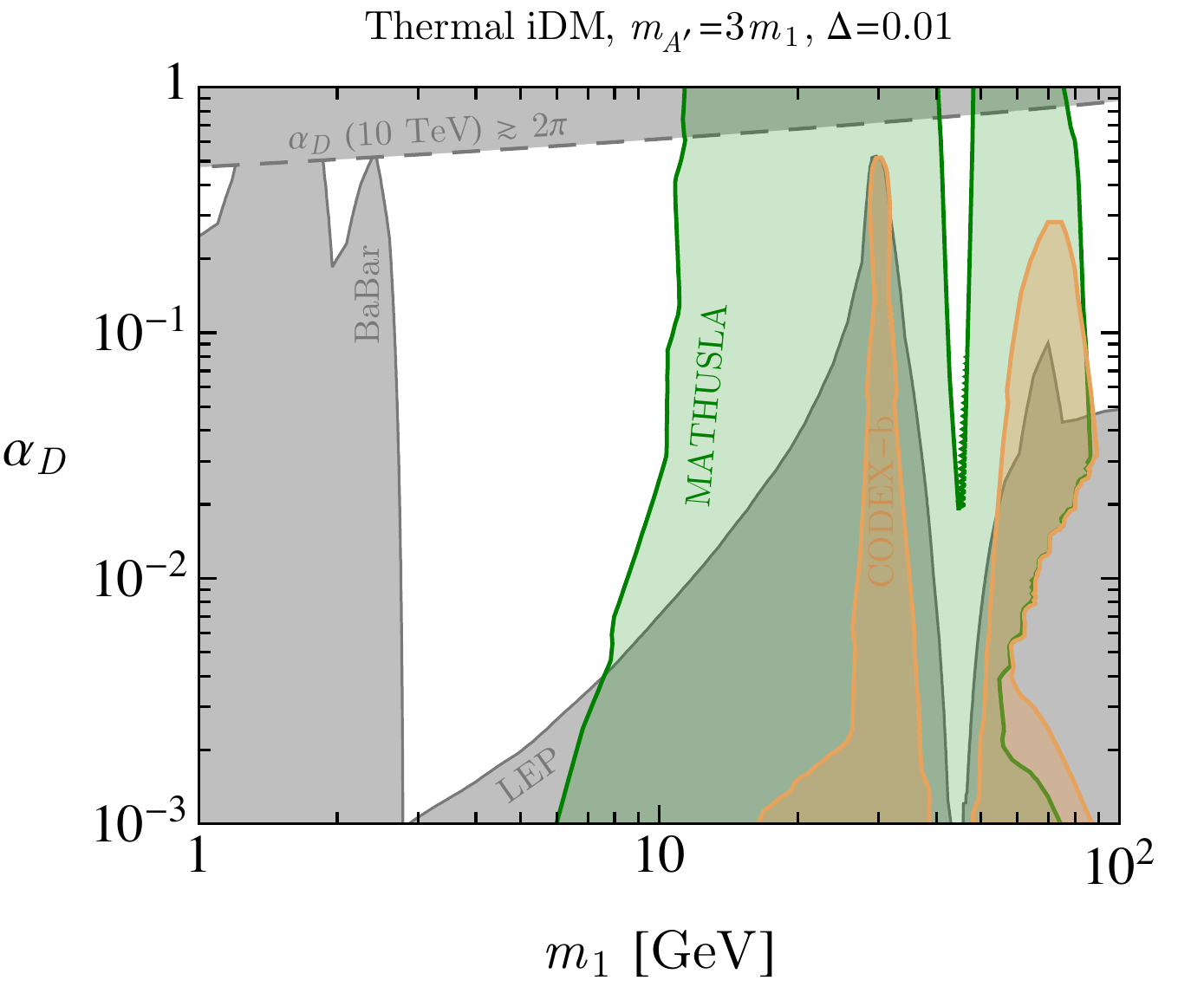}
\caption{As in Fig.~\ref{fig:IDM1}, existing constraints (shaded gray) and projected sensitivities (color) to models of fermionic inelastic dark matter in the $\alpha_D - m_1$ plane for $\mAp / m_1 = 3$ and various choices of the $\x_1 - \x_2$ fractional mass-splitting ($\Delta = 0.1$, 0.05, 0.03, and 0.01). For every point in parameter space, we fix the dark photon kinetic mixing parameter, $\eps$, such that $\x_1$ acquires an adequate cosmological abundance through $\x_1 \x_2 \to A^{\prime *}, Z^* \to f \bar{f}$. As discussed in Sec.~\ref{sec:cosmo}, larger $\alpha_D$ and smaller %$\mAp / m_1$ or 
$\Delta$ correspond to smaller values of $\eps$, suppressing the $\Ap$ production rate at accelerators. Above the gray dashed line, $\alpha_D$ becomes non-perturbative at $\sim 10 \TeV$. The projected sensitivity of Belle II is not shown, since it is nearly identical to that of BaBar in the parameter space shown.}
\label{fig:IDM2}
\end{figure*}
%%%

The right panel of \figref{PvsT} shows similar $\x_2$ kinematic distributions, but now only including events that enter the detector geometry and weighted by the probability to decay within FASER, MATHUSLA, and CODEX-b. Regions of phase space contribute in which $\x_2$ is sufficiently boosted such that the decay length, $\beta\gamma c\tau_{\x_2}$, is not much smaller than the distance between the detector and IP. For the chosen benchmark parameters, FASER is sensitive to the entire available energy spectrum within the angular acceptance, while MATHUSLA and CODEX-b are only able to probe the high-energy tail. 

%%%%%%%%%%%%%%%%%%%%%%%%%%%%%%%%%%%
%%% IDM RESULTS             
%%%%%%%%%%%%%%%%%%%%%%%%%%%%%%%%%%%

%\section{Projected Reach}
\section{Results}
\label{sec:results}

We now discuss the projected sensitivities of the various searches described in \secref{searches}. Our main results are shown in Figs.~\ref{fig:IDM1}-\ref{fig:IDMBig}. In Figs.~\ref{fig:IDM1} and \ref{fig:IDM2}, we focus on the cosmologically motivated  region of $\mAp \sim \text{few} \times m_1 \sim \order{10} \GeV$. 

Fig.~\ref{fig:IDM1} illustrates existing constraints (shaded gray) and the projected sensitivity of proposed LHC searches (solid color) in the $\eps-m_1$ plane for $\mAp / m_1 = 3$, $\alpha_D = 0.1$, and various choices of the fractional DM mass-splitting ($\Delta = 0.1$, 0.05, 0.03, and 0.01). Along the black contour of each panel, the abundance of $\x_1$ agrees with the measured DM energy density. Also shown are the projected reach of the existing Belle II~\cite{Kou:2018nap} and SeaQuest~\cite{Berlin:2018pwi} experiments (dot-dashed). The sensitivity of a monophoton search at Belle II is estimated by rescaling the 20 $\ifb$ background study up to 50 $\iab$~\cite{Inguglia:2016acz,Ferber:2017zjh,DePietro:2018sgg,Kou:2018nap}, and the SeaQuest projection assumes a Phase II luminosity of $10^{20}$ protons on target.

For hidden sector masses slightly above the kinematic threshold of $B$-factories ($\mAp = 3 \, m_1 \gtrsim 10 \GeV$), dedicated LLP searches at CODEX-b (orange), FASER (dark blue), and MATHUSLA (green) are projected to be sensitive to much of the cosmologically motivated parameter space. The right panel of Fig.~\ref{fig:PvsT} aids in understanding the qualitative features of Fig.~\ref{fig:IDM1} and the comparative reach of these experiments. The forward layout of FASER requires the production of highly energetic $\x_1 \x_2$ pairs, enabling the ability to probe shorter proper $\x_2$ lifetimes and hence larger couplings. In contrast, the geometries of CODEX-b and MATHUSLA favor off-axis events that are correspondingly softer. Due to the decreased boost of such events, CODEX-b and MATHUSLA have optimal sensitivity to smaller couplings and longer proper lifetimes. Note that the projected reach of CODEX-b is smaller but qualitatively similar to that of MATHUSLA, due to the former's smaller angular coverage and luminosity. 

%%%
\begin{figure*}[t]
\centering
\includegraphics[width=1\textwidth]{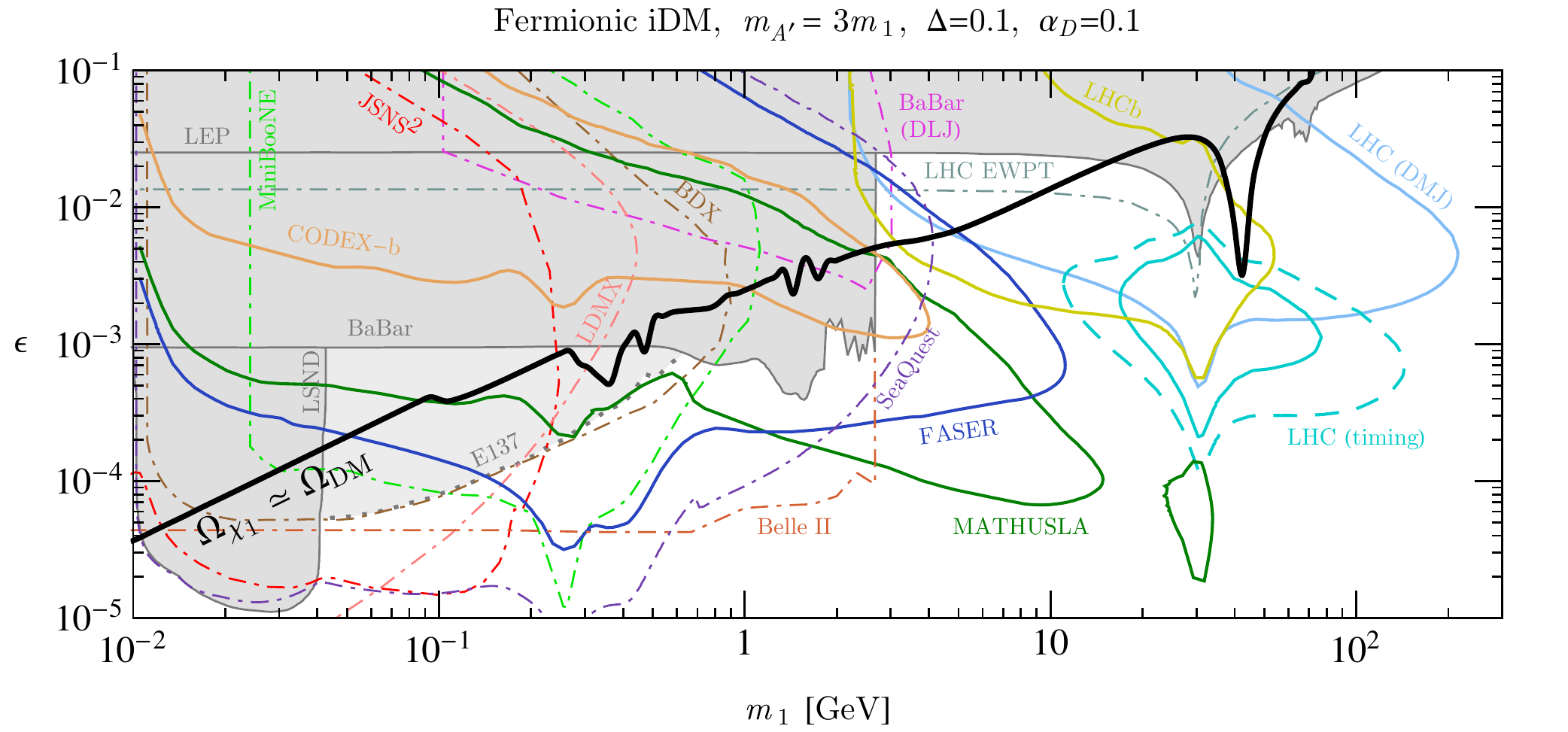}
\caption{As in the top-left panel of Fig.~\ref{fig:IDM1}, existing constraints (shaded gray) and projected sensitivities (color) to models of fermionic inelastic dark matter in the $\eps-m_1$ plane for $\mAp / m_1$ = 3, $\Delta = 0.1$, and $\alpha_D = 0.1$, and now extended to smaller DM masses. Constraints from the SLAC E137 beam dump are shown as a gray dot-dashed contour since they suffer from uncertainties pertaining to the energy threshold of the analysis~\cite{Berlin:2018pwi}. In addition to the various LHC searches discussed in Sec.~\ref{sec:searches}, we also show the projected reach from low-energy accelerators as dot-dashed colored contours, such as Belle II (red)~\cite{Kou:2018nap}, SeaQuest (purple)~\cite{Berlin:2018pwi}, dedicated searches for displaced vertices at BaBar (magenta)~\cite{Izaguirre:2015zva}, MiniBooNE (green)~\cite{Izaguirre:2017bqb}, JSNS$^2$ (red)~\cite{Jordan:2018gcd}, BDX (brown)~\cite{Izaguirre:2017bqb}, and LDMX (pink)~\cite{Berlin:2018pwi,Berlin:2018bsc,Akesson:2018vlm}. Also shown is the future sensitivity from electroweak precision tests (EWPT) at the LHC (dark teal)~\cite{Curtin:2014cca}.}
\label{fig:IDMBig}
\end{figure*}
%%%

Displaced muon searches at LHCb (yellow) and ATLAS/CMS (light blue) are sensitive to smaller displacements and therefore are expected to cover regions of parameter space with larger couplings, provided that the mass-splitting is above the muon threshold ($m_1 > 2 m_\mu / \Delta$). As shown in Fig.~\ref{fig:IDM1}, LHCb is more sensitive to smaller masses (compared to ATLAS/CMS), since in this case the $\x_2$ particles are produced soft and forward and LHCb benefits from its low muon thresholds. In contrast, ATLAS and CMS have enhanced sensitivity to heavier scenarios ($m_{A'} \gtrsim m_Z$), since $\x_2$ is produced more isotropically and its decay products are more energetic.  The proposed LHC timing search utilizing the monojet trigger (solid cyan) extends this reach towards smaller couplings due to the increased size of the accessible decay volume. This search is limited at large couplings by the small $\x_2$ decay length. For smaller masses, $\x_2$ is highly boosted, leading to more collinear decay products and time delays below the required threshold to suppress backgrounds, $\Delta t \lesssim \order{\ns}$.  Also shown in Fig.~\ref{fig:IDM1} is the projected reach of a timing analysis using only a timing-based trigger (dashed cyan), which has enhanced sensitivity compared to one utilizing the conventional monojet trigger. 

Fig.~\ref{fig:IDM2} examines the restricted thermal DM parameter space in the $\alpha_D - m_1$ plane for $\mAp / m_1 = 3$ and various choices of the mass-splitting, $\Delta$. For each panel of this figure, we instead fix $\eps$ to the specific value required for $\x_1$ to freeze out with an abundance that matches the observed DM energy density. Hence, every point in this parameter space is cosmologically viable. In this case, Eq.~(\ref{eq:sigmav1}) implies that for a fixed value of the DM coannihilation cross section at freeze-out, larger values of $\alpha_D$ favor smaller $\eps$, suppressing the $\Ap$ production rate at accelerators. For $m_1 \sim \order{10} \GeV$, the remaining viable parameter space favors sub-10\% mass-splittings ($\Delta \lesssim 0.1$), and sizable $\alpha_D$, i.e., values larger than or comparable to the strength of SM gauge couplings. 

We have ignored the RG-evolution of $\alpha_D$ (e.g., from $m_1 \sim 10 \GeV$ up to LHC energies), which is needed for a more careful comparison between cosmological and accelerator processes. We have bounded $\alpha_D$ from above in Fig.~\ref{fig:IDM2} by demanding that if $\alpha_D$ is defined at $\mu \sim m_1$, then it remains perturbative up to a scale of $\mu \sim 10 \TeV$. This is shown for large $\alpha_D$ as the dashed gray region in Fig.~\ref{fig:IDM2}.

As is evident in Figs.~\ref{fig:IDM1} and \ref{fig:IDM2}, the combined strength of various searches, such as FASER and LHCb, will test most remaining regions of cosmologically viable parameter space for mass-splittings greater than a few percent, while CODEX-b and MATHUSLA will probe thermal relics for even smaller mass-splittings, $\Delta \sim 1\%$. Due to their shorter effective baselines, proposed searches at LHCb and ATLAS/CMS provide the dominant reach for $\Delta \gtrsim 0.1$, while CODEX-b, FASER, and MATHUSLA are increasingly important for DM mass-splittings smaller than $10\%$.

In Figs.~\ref{fig:IDM1} and \ref{fig:IDM2}, we have shown results for $\Delta = 0.1$, 0.05, 0.03, and 0.01. Since the lifetime of $\x_2$ scales as $\tau_{\x_2} \propto \eps^{-2} \Delta^{-5}$ (see Eq.~(\ref{eq:decaychi2V2})), the projected  sensitivities of the various LLP searches shift towards larger values of $\eps$ for smaller mass-splittings. Throughout, we have focused on $\Delta \lesssim 0.1$, since cosmologically motivated values of $\eps$ are often in tension with existing constraints for $m_1 \gtrsim 10 \GeV$ and mass-splittings larger than $\sim 10\%$,  as discussed in Sec.~\ref{sec:cosmo}. Although DM freeze-out in the early universe is largely insensitive to the particular value of $\Delta$ for $\Delta \lesssim 0.1$, the majority of accelerator searches discussed in this work rely on detecting the visible decay products of $\x_2$, whose total energy is directly controlled by the size of $\Delta$. In particular, for $\Delta \lesssim 0.01$, the energy of these decay products often falls under the required experimental thresholds and/or $\x_2$ is so long-lived that it rarely decays inside any of the instruments discussed. Hence, for sub-percent mass-splittings, these accelerator searches do not have significant reach. In this case, future direct detection experiments will have complementary sensitivity for hidden sector gauge couplings near the perturbative limit, i.e., $\alpha_D \gtrsim \order{1}$. For $\Delta \gtrsim 10^{-6}$, inelastic scattering with SM particles is kinematically suppressed, and loop-induced elastic processes dominate. This point is discussed in more detail in Appendix~\ref{sec:DD}.

Although it is not the primary focus of this work, we note that similar signals of iDM have been considered for sub-GeV masses. Fig.~\ref{fig:IDMBig} compares the LHC searches considered here (solid color) to the multitude of previous studies focusing on low-energy accelerators and fixed-target experiments (dot-dashed color) for the standard benchmark of $\mAp / m_1 = 3$, $\Delta = 0.1$, and $\alpha_D = 0.1$. The parameter space shown in Fig.~\ref{fig:IDMBig} is similar to that in the top-left panel of Fig.~\ref{fig:IDM1}, but is now extended down to DM masses of $m_1 = 10 \MeV$. It is interesting to note that in addition to providing leading sensitivity to such models when $m_1 \sim \order{10} \GeV$, proposed LHC experiments such as FASER and MATHUSLA would be competitive with other searches at low-energy accelerators for GeV-scale masses, such as SeaQuest and MiniBooNE. 

In this section, we have fixed the dark photon-to-DM mass ratio to $\mAp / m_1 = 3$. For slightly larger values of $\mAp / m_1$, both the thermal relic line and the projected sensitivities of various LLP searches shift to larger values of $\eps$ compared to Fig.~\ref{fig:IDM1}. We have not considered $\mAp \gg 3 m_1$, since the amount of viable parameter space that is cosmologically motivated shrinks significantly (see Fig.~\ref{fig:IDM_Relic}). For $\mAp \lesssim m_1 + m_2$, the signals discussed above may still occur through the production of off-shell dark photons, although at a suppressed rate. In this case, future LHC searches for visibly decaying dark photons ($\Ap \to \ell^+ \ell^-$) are an efficient means of detection~\cite{Curtin:2014cca}. Furthermore, for $\mAp \lesssim m_1$, there are no longer specific $\eps$-targets in parameter space that are cosmologically motivated, since new processes such as $\x_1 \x_1 \to \Ap \Ap$ facilitate freeze-out in the early universe for much smaller values of $\eps$~\cite{Pospelov:2007mp}. While such scenarios are still interesting and worthy of study, a dedicated analysis is beyond the scope of this work.

%%%
\begin{figure*}[t]
\centering
\includegraphics[width=0.48\textwidth]{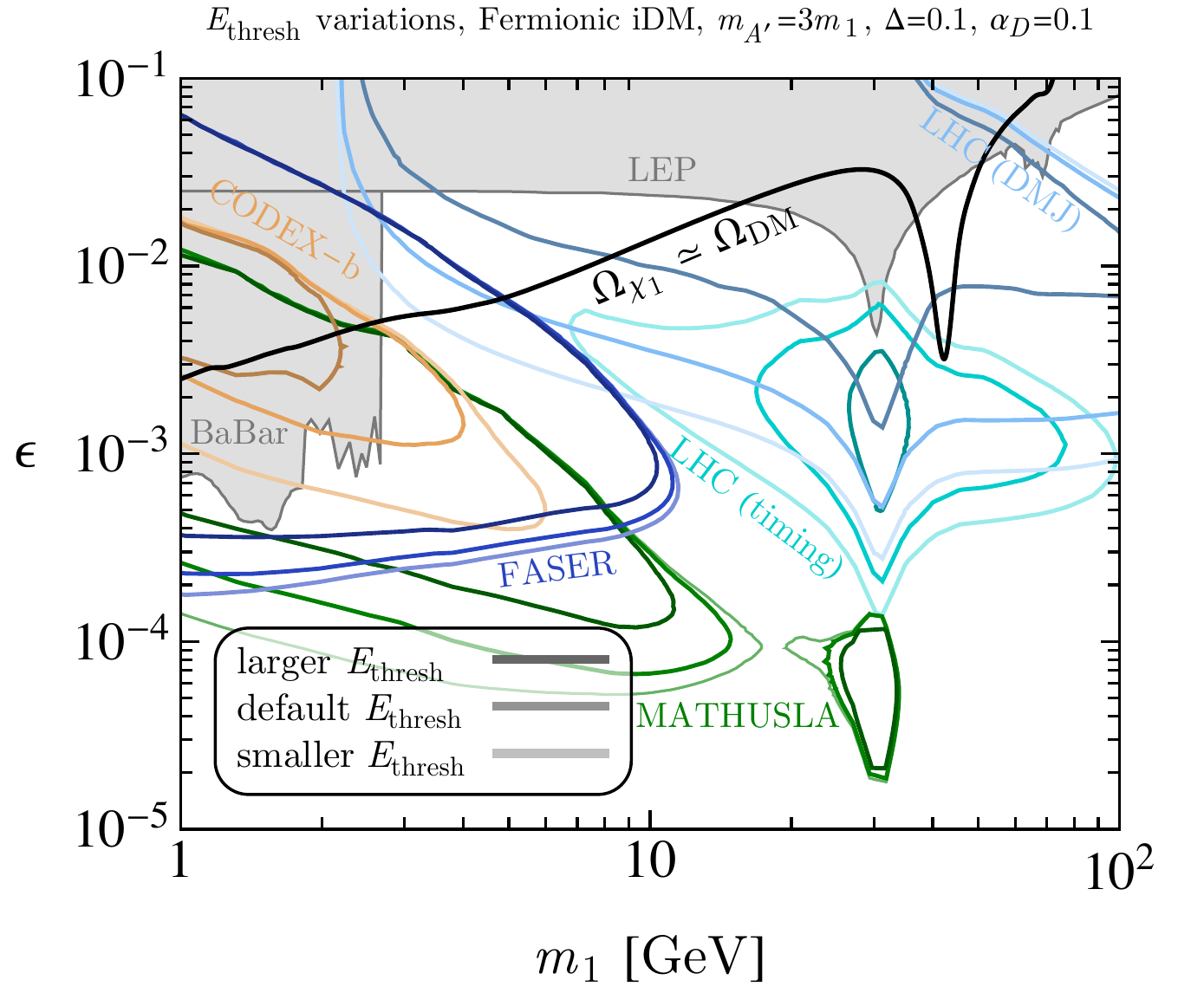}
\includegraphics[width=0.48\textwidth]{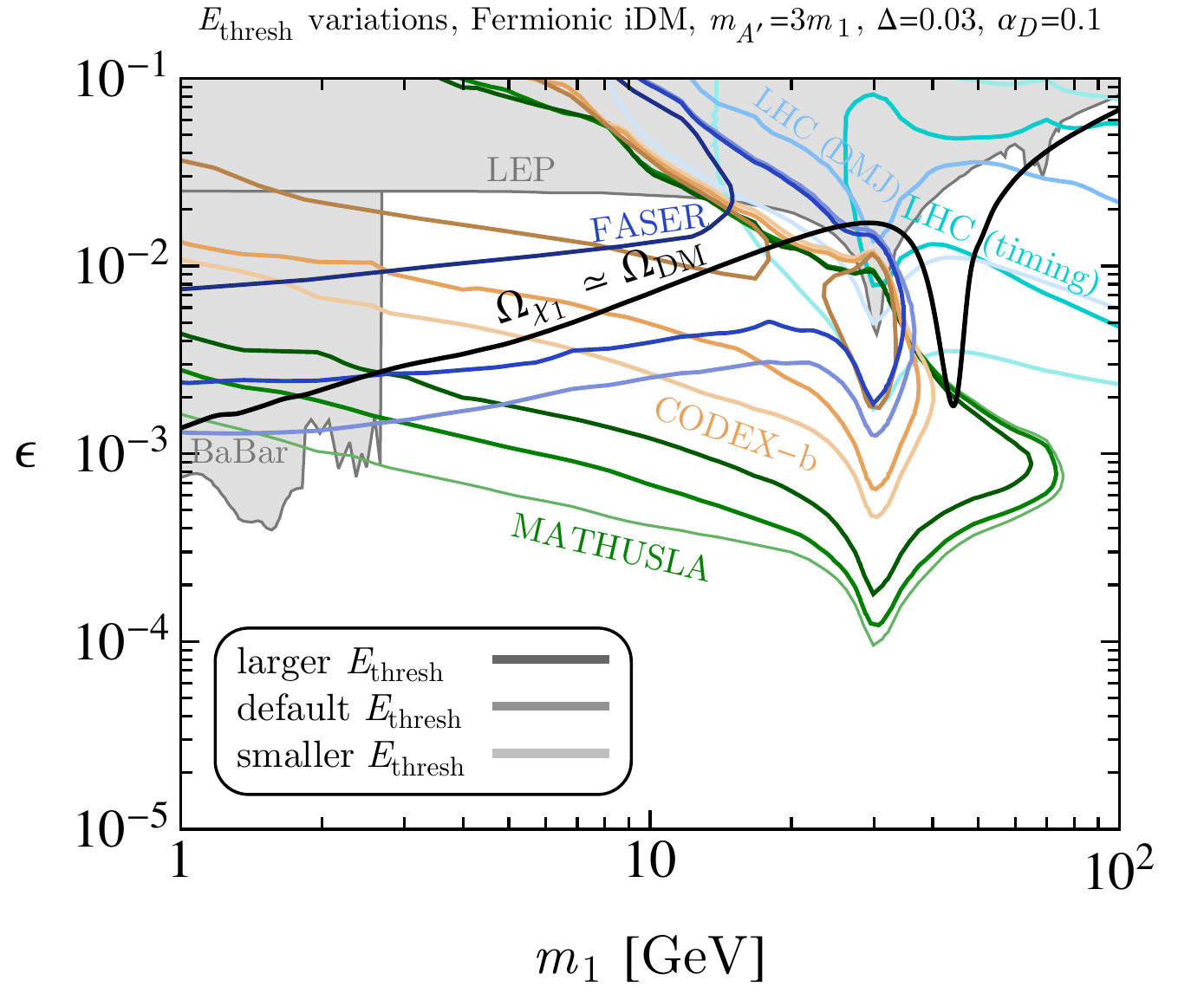}
\caption{As in Fig.~\ref{fig:IDM1}, existing constraints (shaded gray) and projected sensitivities (color) to models of fermionic inelastic dark matter in the $\eps-m_1$ plane for $\mAp / m_1 = 3$, $\alpha_D = 0.1$, and various choices of the $\x_1 - \x_2$ fractional mass-splitting ($\Delta = 0.1$, 0.03), but now including variations of the minimum energy/momentum thresholds for the decay products of $\x_2$. The default thresholds are identical to those shown in Fig.~\ref{fig:IDM1} and detailed in Sec.~\ref{sec:searches}. Darker/lighter shades correspond to larger/smaller minimum energy thresholds for the decay products of $\x_2$. For CODEX-b and MATHUSLA, we demand a minimum energy of 1200, 600, or 300 MeV per track. For FASER, the total visible energy deposition is restricted to be greater than 200, 100, or 50 GeV. For a displaced muon-jet search at ATLAS/CMS and a timing analysis at CMS with a conventional monojet trigger, we implement cuts on the minimum transverse lepton momentum of 10, 5, or 2.5 GeV and 6, 3, or 1.5 GeV, respectively.}
\label{fig:thresh}
\end{figure*}
%%%

%%%%%%%%%%%%%%%%%%%%%%%%%%%%%%%%%%%
%%% Conclusion                           
%%%%%%%%%%%%%%%%%%%%%%%%%%%%%%%%%%%
\section{Conclusion}
\label{sec:conclusion}

GeV-scale hidden sectors have been extensively searched for at low-energy beam dump, fixed-target, and collider experiments. However, many of these searches lack sensitivity if light hidden sectors couple to the Standard Model through mediators  heavier than $\sim 10 \GeV$, as is often appreciated within the context of hidden valleys~\cite{Strassler:2006im}. 

In this work, we have focused on models in which a GeV-scale pseudo-Dirac dark matter multiplet couples to the Standard Model through dark photons of mass $\order{10} \GeV$. In this case, cosmological considerations motivate dark matter fractional mass-splittings smaller than $\sim 10\%$. Furthermore, for mass-splittings larger than $ \order{\text{MeV}}$, the excited dark matter state is often unstable on collider timescales, and future searches for its displaced decay products at existing and proposed LHC experiments can leverage the large center of mass energy to test the majority of remaining motivated parameter space below $\sim 100 \GeV$. 

In this study, we have focused on dedicated searches at ATLAS, CMS, and LHCb, as well as the proposed CODEX-b, FASER, and MATHUSLA experiments. For dark photons heavier than $\sim 10 \GeV$, such experiments constitute the best avenue towards detection. It is enticing to note that even for sub-GeV hidden sector masses, the projected reach of FASER and MATHUSLA are often competitive with higher-intensity and lower-energy beam dump and fixed-target experiments. This warrants more detailed studies of other qualitatively similar scenarios, such as models in which dark matter is the lightest state of a strongly interacting hidden sector~\cite{Hochberg:2014dra,Berlin:2018tvf}.

The visible energy deposition in models of inelastic dark matter depends crucially on the dark matter mass-splitting, and is typically  suppressed for splittings below $\order{100} \MeV$. Hence, unlike, e.g., minimal models of visibly decaying dark photons or dark scalars, enhancing the sensitivity of future LHC searches to models of inelastic dark matter requires optimizing the energy threshold of the analysis to guarantee sufficient signal efficiency while still maintaining adequate background discrimination.  

\section*{Acknowledgements}
%%%%%%%%%%%%%%%%%%%%%%%%%%%%%%%%%%%%%%%%%%%%%%%%%%%%%%
We thank Nikita Blinov, Stefania Gori, Yoni Kahn, Simon Knapen, Gordan Krnjaic, Zhen Liu, Natalia Toro, and Mike Williams for useful discussions. We would also like to personally blame Gordan Krnjaic for his strong work ethic, which is largely responsible for the sheer multitude of contours in Fig.~\ref{fig:IDMBig}. AB is supported by the U.S. Department of Energy under Contract No. DE-AC02-76SF00515. FK is supported by the U.S. National Science Foundation under the grant PHY-1620638. AB is grateful to the hospitality of the Kavli Institute for Theoretical Physics in Santa Barbara, CA, supported in part by the National Science Foundation under Grant No. NSF PHY11-25915 where some of the research reported in this work was carried out. 
%%%%%%%%%%%%%%%%%%%%%%%%%%%%%%%%%%%%%%%%%%%%%%%%%%%%%%

%%%%%%%%%%
\appendix
%%%%%%%%%%%%%%%%%%%%%%%%%%%%%%%%%%%

\section{Energy Thresholds}
\label{sec:threshold}

Fully exploring the cosmologically motivated iDM parameter space with the LHC depends crucially on the energy thresholds implemented in the various analyses of Sec.~\ref{sec:searches}. In this appendix, we briefly discuss how slightly varying these energy/momentum thresholds alters the projected sensitivity of future searches. 

In Fig.~\ref{fig:thresh}, we show the projections of ATLAS, CMS, CODEX-b, FASER, and MATHUSLA  for different energy thresholds, compared to the default baseline cuts of Sec.~\ref{sec:searches}. Darker/lighter shades correspond to larger/smaller minimum energy thresholds for the decay products of $\x_2$. For CODEX-b and MATHUSLA, we demand a minimum energy of 1200, 600, or 300 MeV per track. For FASER, the total visible energy deposition is restricted to be greater than 200, 100, or 50 GeV. For a displaced muon-jet search at ATLAS/CMS we require a minimal muon $p_T$ of 10, 5, or 2.5 GeV. For the timing analysis at CMS, we implement cuts on the minimum transverse lepton momentum of 6, 3, or 1.5 GeV.  As mentioned in Sec.~\ref{sec:searches}, it is beyond the scope of this work to consider such variations for LHCb, although it could significantly enlarge the discovery potential.

Fig.~\ref{fig:thresh} implies that the projected sensitivities are often significantly enhanced for slightly lower energy thresholds. Hence, for sub-10\% DM mass-splittings, the ability to robustly explore regions of cosmology motivated parameter space for a given experimental search depends on the instrument's capability to implement adequately low energy thresholds while still maintaining sufficient background rejection.

%%%%%%%%%%%%%%%%%%%%%%%%%%%%%%%%%%%
%%% Mixing                           
%%%%%%%%%%%%%%%%%%%%%%%%%%%%%%%%%%%
\section{Kinetic Mixing}
\label{sec:mixing}

In this appendix, we briefly review the model of a kinetically-mixed dark photon, closely following the discussion in Ref.~\cite{Curtin:2014cca}. We imagine that at low-energies, the massive gauge boson of a broken $U(1)_D$ symmetry ($\Ap$) kinetically mixes with that of SM hypercharge ($B$). The relevant starting point is
\begin{align}
\label{eq:LagKinMix}
\Lag  &\supset -\frac{1}{4} \, B_{\mu \nu}^2 - \frac{1}{4}  \, A_{\mu \nu}^{\prime \, 2} 
\nl
&~~~
+ \frac{1}{2}  \, \overline{m}_Z^2 \, Z_\mu^2 + \frac{1}{2}  \, \overline{m}_{\Ap}^2 \, A_\mu^{\prime \, 2} 
\nl
&~~~
+ \frac{\eps}{2 \cos{\theta_w}} \, \Ap_{\mu \nu} \, B^{\mu \nu}
~,
\end{align}
where $\theta_w$ is the Weinberg angle and the $Z$ boson is defined as in the SM.  Above, the first two lines contain the relevant kinetic and mass terms for the SM hypercharge, $Z$, and $U(1)_D$ gauge bosons, while the third line introduces kinetic mixing between the two sectors, controlled by the dimensionless parameter, $\eps$. The $Z$ and $\Ap$ mass parameters, $\overline{m}_{Z}$ and $\overline{m}_{\Ap}$, are written in this way to indicate that these are the mass terms in the unmixed basis, and hence correspond to the physical masses only in the limit that $\eps \to 0$. We remain agnostic in regards to the source of $\overline{m}_{\Ap}$, which may arise, for instance, from an extended $U(1)_D$ Higgs sector or the Stueckelberg mechanism. 

Let us first focus on the kinetic terms of Eq.~(\ref{eq:LagKinMix}), which can be diagonalized by performing the following field redefinition on the hypercharge sector,
\be
\label{eq:shift1}
B_\mu \to B_\mu + \frac{\eps}{\cos{\theta_w}} \, \Ap_\mu
~.
\ee
In particular, applying Eq.~(\ref{eq:shift1}) to the kinetic terms of Eq.~(\ref{eq:LagKinMix}) gives
\be
\Lag_\text{kin} = - \frac{1}{4} \, B_{\mu \nu}^2 - \frac{1}{4} \left( 1 - \frac{\eps^2}{\cos^2{\theta_w}} \right) A_{\mu \nu}^{\prime \, 2} 
~.
\ee
In order to canonically normalize the kinetic term of $\Ap$, we perform another redefinition,
\be
\label{eq:shift2}
\Ap_\mu \to \frac{\Ap_\mu}{\left( 1 - \eps^2 / \cos^2{\theta_w} \right)^{1/2}}
~~.
\ee
After this rescaling, we recover the canonical form, i.e.,
\be
\Lag_\text{kin} =  - \frac{1}{4} \, B_{\mu \nu}^2 - \frac{1}{4} A_{\mu \nu}^{\prime \, 2}
~.
\ee

Now, we must examine the mass terms. To recap, Eqs.~(\ref{eq:shift1}) and (\ref{eq:shift2}) can be written as a single transformation,
\begin{align}
\label{eq:shift3}
B_\mu &\to B_\mu + \frac{\eps / \cos{\theta_w}}{\left( 1 - \eps^2 / \cos^2{\theta_w} \right)^{1/2}} ~ \Ap_\mu 
\nl
\Ap_\mu &\to \frac{1}{\left( 1 - \eps^2 / \cos^2{\theta_w} \right)^{1/2}} ~ \Ap_\mu
~,
\end{align}
which allowed us to write the kinetic terms in a canonical form. The first line above corresponds to a shift of the SM $Z$ and photon fields, 
\begin{align}
Z_\mu &\to  Z_\mu - \frac{\eps \, \tan{\theta_w}}{\left( 1 - \eps^2 / \cos^2{\theta_w} \right)^{1/2}} ~ \Ap_\mu
\nl
A_\mu &\to  A_\mu + \frac{\eps}{\left( 1 - \eps^2 / \cos^2{\theta_w} \right)^{1/2}} ~ \Ap_\mu
~.
\end{align}
Since the photon is massless, we only have to consider how the mass terms for the $\Ap$ and $Z$ are affected. In particular, after applying the transformation of Eq.~(\ref{eq:shift3}) to the mass terms of Eq.~(\ref{eq:LagKinMix}), 
the overall mass matrix in the $\left( Z , \Ap \right)$ basis is given by
\be
\label{eq:massmatrix1}
M_{Z \Ap}^2 = \overline{m}_Z^2 ~
\begin{pmatrix}
1 && -\eta
\\  \\ - \eta && \eta^2 + \delta^2
\end{pmatrix}
~.
\ee
where we have defined
\begin{align}
\eta &\equiv \frac{\eps \, \tan{\theta_w}}{\left( 1 - \eps^2 / \cos^2{\theta_w} \right)^{1/2}} 
\nl
\delta &\equiv \frac{\overline{m}_{\Ap} / \overline{m}_Z}{\left( 1 - \eps^2 / \cos^2{\theta_w} \right)^{1/2}} 
~.
\end{align}
Note that the $\Ap-\Ap$  (bottom-right) entry in Eq.~(\ref{eq:massmatrix1}) differs from that of Ref.~\cite{Curtin:2014cca}. To diagonalize this matrix, we need to perform one final field redefinition (which leaves the kinetic terms unchanged),
\be
\label{eq:shift4}
\begin{pmatrix}
Z_\mu \\ \Ap_\mu 
\end{pmatrix}
\to
\begin{pmatrix}
\cos{\alpha} & - \sin{\alpha} \\
 \sin{\alpha} & \cos{\alpha} 
\end{pmatrix}
\begin{pmatrix}
Z_\mu \\ \Ap_\mu 
\end{pmatrix}
~,
\ee
where the angle $\alpha$ is defined as
\begin{align}
\label{eq:alpha}
\tan{\alpha} &= \frac{1}{2 \, \eta} \, \Big[ 1 - \eta^2 - \delta^2 
\nl
&
- \text{sign} (1 - \delta^2) \, \sqrt{ 4 \, \eta^2 + (1 - \eta^2 - \delta^2)^2} \, \Big]
~.
\end{align}
The masses of the $Z$-like and $\Ap$-like states are then given by
\begin{align}
m_{Z, \Ap}^2 &= \frac{\bar{m}_Z^2}{2} \Big( 1 + \eta^2 + \delta^2 
\nl
&\pm \text{sign} (1 - \delta^2) \, \sqrt{(1 +  \eta^2 + \delta^2)^2 - 4 \delta^2} \Big)
~,
\end{align}
where we have adopted the convention of Ref.~\cite{Curtin:2014cca} ($\alpha \to 0$ as $\eps \to 0$ for $m_Z \neq \mAp$, regardless of the sign of $m_Z - \mAp$). It is useful to write down the single overall transformation that combines Eqs.~(\ref{eq:shift3}) and (\ref{eq:shift4}),
\be
\label{eq:basismatrix}
\begin{pmatrix}
\Ap \\ Z \\ A 
\end{pmatrix}
\to
C \times
\begin{pmatrix}
\Ap \\ Z \\ A 
\end{pmatrix}
~,
\ee
where the matrix, $C$, is given by
\be
\label{eq:Cmatrix}
C \equiv
\begin{pmatrix}
(\eta / \eps) \cos{\alpha} \cot{\theta_w} &  (\eta / \eps) \sin{\alpha} \cot{\theta_w} & 0
\\
- (\sin{\alpha} + \eta \cos{\alpha}) &  \cos{\alpha} - \eta \sin{\alpha} & 0
\\
\eta \cos{\alpha} \cot{\theta_w} & \eta \sin{\alpha} \cot{\theta_w} & 1
\end{pmatrix}
.
\ee
This single transformation diagonalizes the entire Lagrangian of Eq.~(\ref{eq:LagKinMix}). 

Let us now apply this transformation to the interaction terms. If the original interaction Lagrangian (in the $\eps \to 0$ limit) is parametrized as
\begin{align}
\label{eq:IntLag0}
\Lag_\text{int} &=  \Ap_\mu \mathcal{J}_D^\mu 
+ \sum\limits_{f} \Big( Z_\mu ~ \bar{f} \gamma^\mu \left( g_v + g_a \gamma^5 \right) f 
\nl
&+ e \, Q_f ~ A_\mu ~ \bar{f} \gamma^\mu f \Big)
~,
\end{align}
then Eq.~(\ref{eq:Cmatrix}) implies that, in the mass basis, we have,
\begin{align}
\label{eq:IntLag1}
&\Lag_\text{int} \to \left( C_{\Ap \Ap} \Ap_\mu  +C_{\Ap Z} Z_\mu \right) \mathcal{J}_D^\mu
\nl
&+ \sum\limits_{f} \Big( \Ap_\mu  \bar{f} \gamma^\mu \left[ (g_v  C_{Z\Ap} + e  Q_f  C_{A\Ap}) + g_a  C_{Z\Ap}  \gamma^5 \right] f
\nl
&+ Z_\mu  \bar{f} \gamma^\mu \left[ (g_v  C_{ZZ} + e  Q_f  C_{AZ}) + g_a  C_{ZZ}  \gamma^5 \right] f \Big)
,
\end{align}
where $\mathcal{J}_D^\mu$ corresponds to the $U(1)_D$ current, the sum is performed over the SM fermions ($f$), and we have ignored couplings involving the massless photon (since it does not couple to $\mathcal{J}_D^\mu$).

It is instructive to take various limits of Eq.~(\ref{eq:Cmatrix}). If $\eps \ll 1$ and $\mAp \ll m_Z$, then Eq.~(\ref{eq:Cmatrix}) becomes
\be
C (\eps \ll 1, \mAp \ll m_Z ) \simeq
\begin{pmatrix}
1 &  - \eps \, \tan{\theta_w} & 0
\\
0 &  1 & 0
\\
\eps & 0 & 1
\end{pmatrix}
.
\ee
In this case, Eq.~(\ref{eq:IntLag1}) is approximately
\begin{align}
&\Lag_\text{int} (\eps \ll 1, \mAp \ll m_Z ) \simeq \left( \Ap_\mu - \eps \, \tan{\theta_w} \, Z_\mu \right) \mathcal{J}_D^\mu 
\nl
&+ \sum\limits_{f} \eps \, e \, Q_f \, \Ap_\mu \, \bar{f} \gamma^\mu f  + \cdots
~,
\end{align}
which is the standard form of the $\Ap-\text{SM}$ interaction that is most often studied in the literature. Another interesting limit arises when $\eps \ll 1$ and $\mAp \simeq m_Z$. In this case, Eq.~(\ref{eq:Cmatrix}) is given by
\be
\label{eq:Cdeg}
C (\eps \ll 1, \mAp \simeq m_Z ) \simeq
\frac{1}{\sqrt{2}} \begin{pmatrix}
1 &  1 & 0
\\
-1 &  1 & 0
\\
\eps & \eps & \sqrt{2}
\end{pmatrix}
.
\ee
It might seem unsettling that the $\Ap-Z$ mixing does not vanish as $\eps \to 0$. However, this is simply a result of the convention for $\alpha$ in Eq.~(\ref{eq:alpha}) when $\mAp \simeq m_Z$. For two degenerate states of identical spin, one can always choose separate bases related by arbitrary orthogonal transformations. Regardless, physical processes must remain unchanged. We will provide a concrete example below.

If we instead choose to parametrize the original interaction Lagrangian of Eq.~(\ref{eq:IntLag0}) as
\be
\Lag_\text{int} =\Ap_\mu \, \mathcal{J}_D^\mu + Z_\mu \, \mathcal{J}_Z^\mu + A_\mu \, \mathcal{J}_\text{em}^\mu
~,
\ee
where $\mathcal{J}_{D, Z, \text{em}}$ are the currents corresponding to the associated gauge bosons, then applying the redefinition of Eq.~(\ref{eq:Cdeg}) gives
\begin{align}
\Lag_\text{int} &\to \frac{1}{\sqrt{2}} ~ \Ap_\mu \left( \mathcal{J}_D^\mu - \mathcal{J}_Z^\mu + \eps \, \mathcal{J}_\text{em}^\mu \right)
\nl
&~
+ \frac{1}{\sqrt{2}} ~ Z_\mu \left( \mathcal{J}_D^\mu + \mathcal{J}_Z^\mu + \eps \, \mathcal{J}_\text{em}^\mu \right) + A_\mu \mathcal{J}_\text{em}^\mu
~.
\end{align}
At energies well below the $\Ap$ and $Z$ masses, we then obtain the corresponding four-Fermi theory coupling the $U(1)_D$ and SM sectors,
\begin{align}
\label{eq:interference}
- \Lag_\text{int} &\simeq \frac{1}{2} \left( \frac{1}{\mAp^2} - \frac{1}{m_Z^2} \right) \, \mathcal{J}_{D \, \mu} \, \mathcal{J}_Z^\mu 
\nl
&
+ \frac{\eps}{2} \left( \frac{1}{\mAp^2} + \frac{1}{m_Z^2} \right) \, \mathcal{J}_{D \, \mu} \, \mathcal{J}_\text{em}^\mu
\nl
& 
\simeq \frac{\eps}{m_Z^2}\, \mathcal{J}_{D \, \mu} \, \mathcal{J}_\text{em}^\mu
~,
\end{align}
which shows that the two sectors indeed decouple for $\eps \to 0$. Note also the lack of a resonant enhancement for $\mAp \simeq m_Z$. 

%%%%%%%%%%%%%%%%%%%%%%%%%%%%%%%%%%%
\section{Direct Detection}
\label{sec:DD}

In this appendix, we provide a calculation for DM elastic scattering in underground direct detection experiments. Let us first briefly summarize the basic formalism. For more comprehensive discussions and reviews, see, e.g., Refs.~\cite{Hill:2014yxa,Hisano:2015bma,Berlin:2015njh}. Let $\x_1$ and $f$ denote a Majorana DM and SM fermion field, respectively. We define the SM currents,
\begin{align}
O_f^{(0)} &\equiv m_f \, \bar{f} f
\nl
O_f^{(2) \mu \nu} &\equiv \frac{1}{2} \, \bar{f} \left[ i \partial^\mu \gamma^\nu + i \partial^\nu \gamma^\mu - \frac{1}{2} g^{\mu \nu} i \slashed{\partial} \right] f
~.
\end{align}
The full theory of Sec.~\ref{sec:idm} is matched to the effective low-energy Lagrangian 
\be
\label{eq:defWilson}
\Lag_\text{EFT} = c_f^{(0)} \, \bar{\x}_1 \x_1  \, O_f^{(0)} + \frac{c_f^{(2)}}{m_1^2} \, \bar{\x}_1 i \partial_\mu i \partial_\nu \x_1  \, O_f^{(2) \mu \nu}
\, ,
\ee
where $c_f^{(0,2)}$ are the spin-0 and spin-2 Wilson coefficients, respectively, and an implicit summation over the SM fermions ($f$) is assumed.

In the case of DM-nucleon scattering ($f=q$, where $q$ is a SM quark), we define the SM form-factors, 
\begin{align}
&\langle N | O_q^{(0)} | N \rangle = m_N ~ f_{q,N}^{(0)}
\nl
& \langle N | O_q^{(2) \mu \nu} | N \rangle = \frac{1}{m_N} \left( k^\mu k^\nu - \frac{1}{4} \, m_N^2 \, g^{\mu \nu} \right) ~ f_{q,N}^{(2)}
~,
\end{align}
where $N = p, n$ denotes a proton or neutron, respectively, and the form-factors, $f_{q,N}^{(0,2)}$, are given by the lattice and nucleon PDFs, respectively~\cite{Hill:2014yxa,Hisano:2015bma,Berlin:2015njh} . This can be adapted in the case of electron-scattering ($f=e$), in which case we have
\begin{align}
&\langle e | O_e^{(0)} | e \rangle = m_e
\nl
& \langle e | O_e^{(2) \mu \nu} | e \rangle = \frac{1}{m_e} \left( k^\mu k^\nu - \frac{1}{4} \, m_e^2 \, g^{\mu \nu} \right)
~.
\end{align}

These matrix elements, together with the Wilson coefficients of Eq.~(\ref{eq:defWilson}), define the spin-independent amplitudes for nucleon and electron scattering 
\begin{align}
\label{eq:DDAmp}
\mathcal{M}_N &= m_N \bigg( \sum\limits_{q=u,d,s} f_{q,N}^{(0)} \, c_q^{(0)} + \frac{3}{4} \, f_{q,N}^{(2)} \, c_q^{(2)} \bigg)
\nl
\mathcal{M}_e &= m_e \bigg( c_e^{(0)} + \frac{3}{4} \, c_e^{(2)} \bigg)
~.
\end{align}
We then have for the spin-independent scattering cross section,
\be
\sigma_\text{SI} = \frac{4}{\pi} ~ \mu_{1 i}^2 ~ |\mathcal{M}_i|^2
~,
\ee
where $i = N, e$ and $\mu_{1 i}$ is the $\x_1 - i$ reduced mass.

%%%
\begin{figure}[t]
\centering
\includegraphics[width=0.3\textwidth]{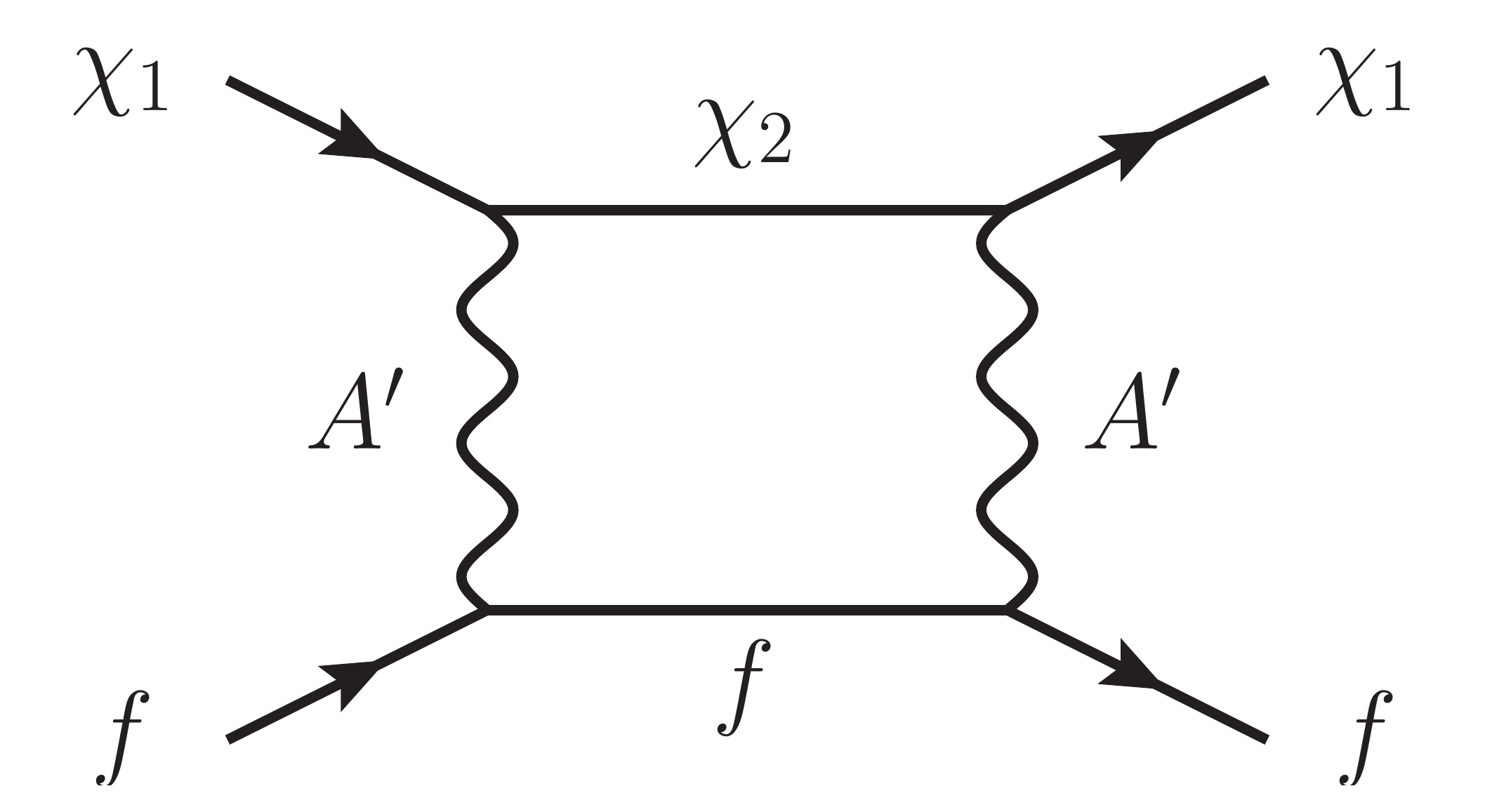}
\caption{A representative Feynman diagram responsible for the one-loop coupling of DM ($\x_1$) to SM fermions ($f$) (crossed diagrams not shown). There are also one-loop $Z$-exchange diagrams and two-loop diagrams couplings $\x_1$ to the gluon field-strength. We do not include these other contributions in our estimate for the DM elastic scattering cross section.}
\label{fig:box}
\end{figure}
%%%

For the model of fermionic iDM in Sec.~\ref{sec:idm}, inelastic scattering ($\x_1 f \to \x_2 f$) is kinematically suppressed for $\Delta \gtrsim 10^{-6}$. In this case, the leading order contribution to elastic scattering is given by one-loop diagrams, as shown in Fig.~\ref{fig:box}. Similar diagrams exist for $Z$-exchange, and $\x_1$ also couples to the gluon field-strength at the two-loop level. A detailed calculation is beyond the scope of this work, but we do not expect these contributions to significantly alter our final results. 

In the limit that $m_f \ll m_1 \ll \mAp$, we calculate the Wilson coefficients from the diagram of Fig.~\ref{fig:box} to be 
\begin{align}
c_f^{(0)} &\simeq - \, \frac{3 \, Q_f^2 \, \alpha_\text{em} \, \alpha_D \, \eps^2 \, m_1}{2 \, m_{A^\prime}^4} ~ \left( 1 - 4 \, \log{\frac{m_{A^\prime}}{m_1}} \right)
\nl
c_f^{(2)} &\simeq - \, \frac{4 \, Q_f^2 \, \alpha_\text{em} \, \alpha_D \, \eps^2 \, m_1}{9 \, m_{A^\prime}^4} ~ \left( 1 - 12 \, \log{\frac{m_{A^\prime}}{m_1}} \right)
~.
\end{align}
In the case of $\x_1$-electron scattering, this gives a simple analytic result for the cross section,
\be
\label{eq:DDelec}
\sigma_\text{SI}^\text{(electron)} \simeq \frac{\alpha_\text{em}^2 \, \alpha_D^2 \, \eps^4 \, m_e^4 \, m_1^2}{9 \pi \, m_{A^\prime}^8} ~ \left( 11 - 60 \, \log{\frac{m_{A^\prime}}{m_1}} \right)^2
~.
\ee
Although a detailed calculation for the DM-nucleon cross section should include the appropriate form-factors, as in the top line of Eq.~(\ref{eq:DDAmp}), the form of Eq.~(\ref{eq:DDelec}) is still useful as a parametric estimate,
\be
\label{eq:DDnucleon}
\sigma_\text{SI}^\text{(nucleon)} \sim \frac{\alpha_\text{em}^2 \, \alpha_D^2 \, \eps^4 \, m_N^4 \, m_1^2}{m_{A^\prime}^8}
~.
\ee
We demand that the DM-nucleon elastic scattering cross section is larger than the coherent neutrino background, i.e., $\sigma_\text{SI}^\text{(nucleon)} \gtrsim 10^{-49} \text{ cm}^2 \times (m_1 / 10 \GeV)$~\cite{Billard:2013qya}. Eq.~(\ref{eq:DDnucleon}) implies that this is achieved when
\be
\alpha_D \, \eps^2 \gtrsim 10^{-4} \times \left( \frac{m_1}{10 \GeV} \right)^{7/2} \left( \frac{\mAp}{3 \, m_1} \right)^4
~.
\ee
Hence, future experiments such as LUX~\cite{Akerib:2018lyp} will only be competitive with existing constraints from LEP~\cite{Hook:2010tw,Curtin:2014cca} for $\alpha_D \gtrsim \order{1} \times (\mAp / 3 \, m_1)^4$.

%%%%%%%%%%%%%%%%%%%%%%%%%%%%%%%%%%%
\section{Dark Photon Mixing with Vector Mesons}
\label{sec:VectorMesonMixing}

According to the model of vector meson dominance (VMD), the photon couples to hadronic states through mixing with intermediate QCD vector mesons.\footnote{An overview of VMD is given in, e.g., Ref.~\cite{OConnell:1995nse}. For a comparison between different conventions, see Ref.~\cite{Zerwekh:2006tg}.} Following the convention in which the SM photon ($A^\mu$) and vector mesons ($V^\mu$) mass-mix, the corresponding effective Lagrangian is given by
\be
\label{eq:VMD1}
- \Lag \subset (e \, m_V^2 / g_V) ~ A^\mu \, V_\mu
\, ,
\ee
where $g_V$ is the vector meson-pion interaction strength.
If a light dark photon ($m_{A'} \ll m_Z$) kinetically mixes with SM hypercharge, the photon field is redefined as $A_\mu \rightarrow A_\mu + \eps A'_\mu$ (see Appendix~\ref{sec:mixing}). This leads to an effective mass-mixing between the dark photon and vector mesons that is analogous to Eq.~(\ref{eq:VMD1}),
\be
\label{eq:VMD2}
- \Lag \subset (\eps \, e \, m_V^2 / g_V) ~ A^{\prime \mu} \, V_\mu
\ee

The amplitudes for dark photon and vector meson production are related by the $\Ap-V$ mixing of Eq.~(\ref{eq:VMD2}). Let us denote the amplitude to produce a vector meson as $\mathcal{M}_V$ and the amplitude to produce a dark photon as $\mathcal{M}_{A'}$. We can then write
\be
\label{eq:VMD3}
\mathcal{M}_{A'} \simeq \theta_{V} \, \mathcal{M}_V
\, ,
\ee
where the mixing parameter, $\theta_V$, is defined as 
\be
 \theta_V \equiv  \frac{\eps \, e }{g_V} ~ \frac{m_V^2}{q^2 - m_V^2 + i m_V \Gamma_V}  
 ~ ,
\ee
$\Gamma_V$ is the vector meson width, and $q^2 = \mAp^2$ corresponds to on-shell $\Ap$ production. 

There are multiple vector mesons that mix with $\Ap$ and contribute to its production, such as the $\rho$, $\omega$, and heavier hadronic resonances. In this study, we only consider the leading contribution from $\Ap-\rho$ mixing and neglect possible contributions from other vector mesons. This is motivated by the fact that the production rate of heavier excited states and $\Ap-\omega$ mixing are suppressed in comparison, e.g., $\theta_{\omega}/\theta_\rho \sim g_\rho/g_\omega \simeq 0.3$ where $g_\rho\simeq 5$ and $g_\omega \simeq 17$
~\cite{Dubnicka:2002yp}. 

Eq.~(\ref{eq:VMD3}) allows us to obtain an approximate $\Ap$ spectrum by reweighting that of the $\rho$ meson, i.e., $d \sigma_{A'} = \theta_\rho^2 ~ d \sigma_\rho$. In this work, we use the inclusive $\rho$ meson spectrum provided by the Monte Carlo generator EPOS-LHC~\cite{Pierog:2013ria}. We have also checked that this procedure qualitatively matches forward Bremsstrahlung production of dark photons, as discussed in Sec.~\ref{sec:prod}.

While the above method can be generalized to include additional mesons, this would require knowledge of the specific vector meson production processes in order to incorporate interference effects, which are known to be important in modeling the suppressed $\Ap$ production rate for $\mAp \gtrsim \text{GeV}$~\cite{Blumlein:2013cua,Gorbunov:2014wqa,deNiverville:2016rqh}. In general, this is only available in special cases such as  Bremsstrahlung, in which the dark photon is assumed to mix with vector mesons that are radiated off of the incoming proton beam. 

%++++++++++++++++++++++++++++++++++++++++++++++++++++++++
%References
%++++++++++++++++++++++++++++++++++++++++++++++++++++++++

\bibliographystyle{utphys}
\bibliography{idm_faser}

\end{document}